\documentclass{article}

\usepackage{arxiv}

\usepackage{amsmath,amssymb,amsfonts}

\usepackage[utf8]{inputenc} 
\usepackage[T1]{fontenc}    
\usepackage{hyperref}       
\usepackage{url}            
\usepackage{booktabs}       
\usepackage{amsfonts}       
\usepackage{nicefrac}       
\usepackage{microtype}      
\usepackage{lipsum}         
\usepackage{graphicx}
\usepackage{natbib}
\usepackage{doi}

\usepackage{algorithm}
\usepackage{algorithmicx}
\usepackage{algpseudocode}

\usepackage{listings}
\usepackage{textcomp}
\lstdefinestyle{cppstyle}{
  language=C++,
  basicstyle=\ttfamily\footnotesize,
  morekeywords={uint32_t, __half},
}
\lstdefinestyle{cppstyle-inline}{
  language=C++,
  basicstyle=\ttfamily,
}
\usepackage{xcolor}
\usepackage{multirow}
\usepackage[export]{adjustbox}
\usepackage{makecell}
{}

\newcommand{\safeincludegraphics}[3]{%
  \IfFileExists{#3}{%
    \includegraphics[#1]{#3}%
  }{%
    \includegraphics[draft,#1,#2]{#3}%
  }%
}

\newcommand{\mname}[1]{%
  {\ttfamily\detokenize{#1}}%
}

\usepackage{siunitx}
\usepackage{threeparttable}

\usepackage[caption=false,font=footnotesize]{subfig}

\usepackage[nameinlink]{cleveref}
\crefname{equation}{}{}
\Crefname{equation}{Equation}{Equations}
\crefname{lstlisting}{listing}{listings}
\Crefname{lstlisting}{Listing}{Listings}
\crefname{figure}{Figure}{Figures}
\Crefname{figure}{Figure}{Figures}
\crefname{subfigure}{Figure}{Figures}
\Crefname{subfigure}{Figure}{Figures}
\crefname{table}{Table}{Tables}
\Crefname{table}{Table}{Tables}
\crefname{section}{Section}{Sections}
\Crefname{section}{Section}{Sections}

\usepackage{comment}

\newcommand{\proposed}{PackSELL}
\newcommand{\suitable}{SELL-suitable}

\newcommand{\ieee}{IEEE 754}
\newcommand{\abbrev}[1]{\textbf{#1}}

\begin{document}

\title{\proposed{}: A Sparse Matrix Format for Precision-Agnostic High-Performance SpMV}

\author{ \href{https://orcid.org/0009-0002-8328-0388}{\includegraphics[scale=0.06]{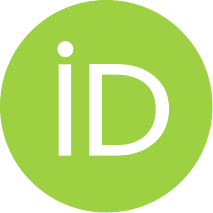}\hspace{1mm}Kengo~Suzuki} \\
    Academic Center for Computing and Media Studies, \\
    Kyoto University, \\
    Kyoto, Japan \\
    \texttt{suzuki@i.kyoto-u.ac.jp} \\
    \And
    Takeshi~Iwashita \\
    Academic Center for Computing and Media Studies, \\
    Kyoto University, \\
    Kyoto, Japan \\
    \texttt{iwashita@i.kyoto-u.ac.jp} \\
}

\renewcommand{\shorttitle}{\proposed{}: A Sparse Matrix Format for Precision-Agnostic High-Performance SpMV}

\maketitle

\begin{abstract}
  We propose a new sparse matrix format, \proposed{}, designed to support diverse data representations and enable efficient sparse matrix–vector multiplication (SpMV) on GPUs.
  Building on sliced ELLPACK (SELL), \proposed{} incorporates delta encoding of column indices and a novel packing scheme that stores each index-delta–value pair in a single word, thereby reducing memory footprint and data movement.
  This design further enables fine-grained control over the bit allocation between deltas and values, allowing flexible data representations, including non-IEEE formats.
  Experimental results show that, when configured for half precision (FP16), the \proposed{}-based SpMV kernel outperforms the cuSPARSE SELL-based kernel by up to 1.63$\times$.
  Moreover, with configurations using customized formats, \proposed{} achieves FP32-level accuracy while exceeding the performance of FP16 cuSPARSE.
  These benefits extend to sparse linear solvers;
  for example, a mixed-precision preconditioned conjugate gradient (PCG) solver using \proposed{} achieves up to a 2.09$\times$ speedup over the standard full-precision PCG.
\end{abstract}


\section{Introduction}
Sparse matrix-vector multiplication (\abbrev{SpMV}) is a fundamental kernel in a wide range of scientific applications and has been extensively studied~\citep{gao2024SystematicLiterature}.
In this study, we consider the efficient execution of SpMV on GPUs for real matrices,
\begin{equation}
  \label{eq:spmv}
  y = Ax,\,\, A = [a_{i,j}] \in \mathbb{R}^{n\times m},\,\, x \in \mathbb{R}^{m},\,\,y \in \mathbb{R}^{n},
\end{equation}
for different low-precision data representations.
In particular, we develop a new sparse matrix format that efficiently and uniformly supports various data representations, including integer and floating-point formats not defined in \ieee{}.

Recent hardware trends have widened the gap between memory bandwidth and computing performance, especially with advancements in low-precision arithmetic units~\citep{lindquistReducingCommunication,abdelfattah2021SurveyNumerical}.
As a result, reducing data movement has become critical for memory-bound kernels such as SpMV.
Consequently, mixed-precision techniques, particularly those using low precision, have attracted attention~\citep{abdelfattah2021SurveyNumerical}.
A prominent example is iterative solvers for sparse linear systems, where the \ieee{} single-precision (\abbrev{FP32}) format has become widely used in addition to the double-precision (\abbrev{FP64}) format~\citep{buttari2008UsingMixed,ikuno2012IterativeSolver,nakajima2021EfficientParallel,lindquist2022AcceleratingRestarted,zhao2022NumericalInvestigation,zhao2023NumericalBehavior,amestoy2024FivePrecisionGMRESBased,spyropoulos2025Numericalstudy}.
More recently, state-of-the-art methods exploit the half-precision floating-point (\abbrev{FP16}) format~\citep{suzuki2025NestedKrylov}.
Furthermore, non-IEEE floating-point~\citep{ichimura2018FastScalable,hunhold2025EvaluationBfloat16} and few-bit integer (or fixed-point)~\citep{jerez2015LowComplexity,iwashita2020integerarithmeticbased,suzuki2025IntegerArithmeticBased} representations have also been investigated.

These trends suggest the growing significance of supporting various low-bit formats in memory-bound kernels, including SpMV.
However, existing high-performance SpMV kernels are still largely designed for classical higher-precision IEEE formats such as FP64 and FP32, and the investigation of other formats remains limited.
This is mainly due to the tight coupling between data representation and the efficiency of memory access:
conventional sparse matrix formats implicitly assume data types aligned to hardware-friendly boundaries, typically multiples of bytes.
Although several custom formats are designed using memory accessors to better utilize memory bandwidth~\citep{anderson2016VectorizationMultibyte,mukunoki2023SparseMatrixVector,graillat2024AdaptivePrecision}, they are still constrained by such alignment requirements, and the variability of available data representations is limited.
Consequently, many potentially useful non-byte-aligned formats remain difficult to use efficiently.

We argue that this limitation stems from treating data representation as a fixed attribute when designing sparse matrix formats.
That is, the limitation can be overcome by jointly designing data representation together with other attributes of sparse matrix formats, such as index encoding and memory layout.
To this end, we propose a new sparse matrix format based on the Sliced ELLPACK (\abbrev{SELL}) format~\citep{monakov2010AutomaticallyTuninga}, called delta–value packing-based SELL (\abbrev{\proposed{}}).
The proposed format packs each nonzero element together with its column index encoded as the difference (delta) from the preceding element into a single word;
these packed data are reconstructed on the fly during operations such as SpMV.
By flexibly adjusting the bit allocation between deltas and nonzero values, \proposed{} supports a wide range of data representations, including non-byte-aligned formats, while reducing memory footprint through delta encoding and packing.
Therefore, \proposed{} is expected to enable both flexible data representations and high-performance SpMV.

We evaluate a \proposed{}-based GPU SpMV kernel across a range of bit allocations between values and deltas.
In an FP16-equivalent setting, our method outperforms conventional SpMV kernels, including NVIDIA cuSPARSE and the state-of-the-art DASP~\citep{lu2023DASPSpecific}.
We further examine the trade-off between performance and accuracy by varying the bit allocation, showing that FP32-level accuracy can be achieved with FP16-level performance.
Finally, we integrate our kernel into mixed-precision Krylov subspace methods and demonstrate overall speedups and a practical scenario where non-IEEE formats realized by \proposed{} provide performance benefits.

The contributions of this study are summarized as follows:
\begin{itemize}
  \item A novel sparse matrix format, \proposed{}, enabling high-performance SpMV on GPUs across various data formats.
  \item Performance analysis of SpMV kernels and Krylov subspace methods using \proposed{}.
  \item Identification of a scenario where non-IEEE representations benefit mixed-precision Krylov subspace methods.
\end{itemize}

\section{Related Work} \label{sec:related_work}
Since it is practical, SpMV has been researched on various architectures, including GPUs~\citep{filippone2017SparseMatrixVector,gao2024SystematicLiterature}.
Previous studies can be broadly classified into two approaches: improving execution efficiency (e.g., load balancing and memory access optimization) and reducing memory footprint via data compression, although some studies address both simultaneously~\citep{aliaga2022CompressionLoad}.

Both approaches are effective, but they generally target different types of matrices.
The former focuses on matrices with irregular sparsity patterns, where workload imbalance is a major bottleneck, especially on GPU SIMT architecture.
To address this, formats such as compressed sparse row (\abbrev{CSR}) and coordinate (\abbrev{COO}), as well as their variants, are widely adopted, and numerous optimization techniques have been proposed~\citep{kreutzer2014unifiedsparse,anzt2014ImplementingSparse,bell2009Implementingsparse,ashari2014FastSparse,liu2015CSR5Efficient,steinberger2017Globallyhomogeneous,anzt2020LoadbalancingSparse,niu2021TileSpMVTiled}.
Adaptive techniques that hybridize multiple sparse matrix formats or select multiple sub-kernels based on matrix characteristics have further improved performance~\citep{lu2023DASPSpecific}.

However, these strategies often provide limited benefits for matrices with relatively regular sparsity patterns, as suggested in~\citep{zhang2025CanTensor}.
For such matrices, straightforward implementations using hardware-friendly formats such as SELL can already achieve near-peak performance, as bounded by memory bandwidth.
In this case, reducing memory footprint is critical to boost performance.
Moreover, since such matrices typically arise in applications such as sparse linear solvers, this direction is also important in practice.

One strategy in this direction is to compress indices.
Traditional techniques include blocking methods, such as block sparse row (\abbrev{BSR})~\citep{barrett1994Templatessolution} and block-based SELL variants, which reduce index size by grouping nonzero elements into blocks.
While effective for structured matrices, these methods can introduce excessive padding and degrade performance for other matrices.
Column-wise blocking has also been proposed for a similar purpose~\citep{nagasaka2016AdaptiveMultilevel}, and recent work continues to refine such strategies~\citep{cong2025CBSpMVData}.
Other techniques compress indices, for example, using dictionary compression~\citep{murakami2026CoDSELLNonZero} or encoding differences between neighboring elements~\citep{maggioni2014CoAdELLAdaptivity}.
Although the latter is closely related to our delta encoding, these methods treat index and value representations separately.

In addition to index compression, several studies have explored compressing matrix values.
Assuming many elements have similar values, lossless techniques such as delta encoding and run-length encoding have been applied to nonzero elements~\citep{wolfgang2024ValueCompressedSparse,galanopoulos2025DIVIndex}.
However, these approaches primarily target CPUs, and efficient GPU implementations remain challenging.

With the growing adoption of mixed-precision algorithms, lossy approaches based on reduced precision have also gained attention.
Simplified strategies use lower-precision IEEE~754 formats (FP32 and FP16), while advanced ones employ non-IEEE representations with custom memory accessors~\citep{kawai2022LowAdaptive,mukunoki2023SparseMatrixVector,graillat2024ReducedPrecisionReducedExponent}.
Some approaches further combine multiple data types within a matrix based on error estimation~\citep{graillat2024AdaptivePrecision}.
While these techniques can effectively reduce memory footprint when accuracy requirements permit, they often require byte-aligned formats, introduce padding to ensure alignment, or impose a constraint on the number of elements processed per memory access.
These constraints limit the variety of data representations and GPU efficiency.

Overall, previous methods improve SpMV performance for matrices with regular sparsity patterns primarily through data reduction.
However, these methods typically treat values and indices separately, imposing alignment constraints on their individual storage and limiting the flexibility in data representation.
By contrast, this study proposes a sparse matrix format that jointly compresses matrix values and indices for such matrices.
By integrating the design of data representation, index encoding, and memory layout, the proposed format enables efficient GPU execution and flexible data representations, including non-byte-aligned formats.

\section{SELL format} \label{sec:sell}
This section reviews the SELL format~\citep{monakov2010AutomaticallyTuninga}, the basis of the proposed \proposed{} format.
SELL is a state-of-the-art sparse matrix format that groups consecutive rows into slices and aligns the nonzero elements within each slice.
This alignment enables a uniform workload and efficient memory access, making SELL well suited for matrices with relatively regular sparsity patterns on SIMD and SIMT architectures.

\begin{figure}
  \centering
  \includegraphics[width=0.6\linewidth]{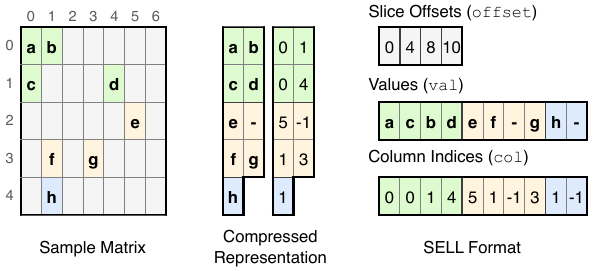}
  \caption{Example of the SELL format with slice size $C = 2$.}
  \label{fig:sell}
\end{figure}

Specifically, as shown in \cref{fig:sell}, SELL represents a sparse matrix using three arrays, denoted here as \texttt{val}, \texttt{col}, and \texttt{offset}.
The arrays \texttt{val} and \texttt{col} store the nonzero values and their corresponding column indices, respectively, in column-major order.
However, rows are partitioned into slices of consecutive $C$ rows, and padding values may also be stored so that all rows within a slice have the same number of elements.
This padding ensures that all rows in a slice have the same computational workload, which facilitates SIMD and SIMT execution.
The starting positions of the slices in \texttt{val} and \texttt{col} are managed in the \texttt{offset} array.

When nearby rows have significantly different numbers of nonzero elements, SELL may introduce substantial padding, particularly for large values of $C$.
To mitigate this issue, SELL is often combined with row permutation, where rows are sorted in descending order by the number of nonzero elements per row within blocks of $\sigma$ rows.
This strategy increases the likelihood that rows within each slice have similar numbers of nonzero elements, reducing padding.
This technique is known as SELL-$C$-$\sigma$~\citep{kreutzer2014unifiedsparse}.

Depending on the application, the row permutation can be applied either explicitly or implicitly.
If the correctness of the algorithm is unaffected by the row permutation, explicitly reordering the matrix is typically preferable, as it simplifies the implementation and subsequent computations.
In this case, the reordered matrix is handled directly using the standard SELL format.
Conversely, if the original ordering must be preserved, reordering should be applied implicitly during storage and reverted during computation.

A straightforward way to implement implicit reordering is to introduce an additional unsigned integer array, \texttt{perm}, that stores permutation information.
While several designs are possible, in this study, we store permutation data for every $\sigma$ rows to save memory usage;
specifically, the array stores values in the range 0 to $\sigma-1$ corresponding to the permutation within each block of $\sigma$ rows.
The data type is chosen to be the minimum required to represent $\sigma-1$; for example, an 8-bit unsigned integer suffices when $\sigma \le 256$.

In this implicit case, a simple SpMV algorithm using SELL-$C$-$\sigma$ is given below; it employs 0-based indexing:
\begin{algorithmic}[1]
  \For{$i = 0,\ldots,n-1$}
  \State $i' := \lfloor i / \sigma \rfloor \cdot \sigma + \texttt{perm}_i$
  \State $k := \lfloor i / C \rfloor$
  \State $l := i \bmod C$
  \State $s := \texttt{offset}_k$
  \State $w := (\texttt{offset}_{k + 1} - s) / C$
  \State $t := 0$
  \For{$j=0,\ldots,w-1$}
  \State $p := s + j \cdot C + l$
  \State $t := t + \texttt{val}_p \cdot x_{\texttt{col}_p}$
  \EndFor
  \State $y_{i'} := t$
  \EndFor
\end{algorithmic}
Here, a subscripted array name, such as $\texttt{val}_p$, denotes the element at the corresponding index.

\section{\proposed{} format}
SELL-based SpMV performs particularly well for large-scale matrices with relatively regular sparsity patterns, often achieving performance close to the memory bandwidth limit.
In such cases, reducing data movement becomes a key direction for further improving performance.
In addition, such matrices frequently arise in scientific simulations, where mixed-precision (low-precision) algorithms are actively studied.
Motivated by these observations, we propose a new sparse matrix format, \proposed{}, built upon the SELL format.

Like SELL, \proposed{} considers slices of size $C$ and stores nonzero elements with padding so that all rows within each slice have the same number of elements, and in turn, the same workload.
To further reduce memory footprint and increase the flexibility of data representation, \proposed{} introduces two key techniques: delta encoding for column indices and packing each pair of an index delta and a nonzero value into a single word.
Both contribute to reduced memory footprint, and the packing technique provides flexible data representations, covering even non-\ieee{} formats.

\subsection{Delta Encoding for Column Indices}
The first key feature is delta encoding of column indices, in which the column index of each nonzero element is represented as the difference from the preceding nonzero element in the same row.
Formally, for the $i$-th row, the column index of a nonzero element $a_{i, j}$ is encoded as a non-negative delta $d_{i,j}$:
\begin{equation}
  d_{i,j} =
  \begin{cases}
    j - \max S_{i,j}   & \text{if } S_{i,j} := \{k<j \mid a_{i,k} \neq 0\} \neq \varnothing, \\
    j - \mathfrak{d}_i & \text{otherwise},
  \end{cases}
\end{equation}
where $\mathfrak{d}_i$ is an offset used for the leftmost nonzero element, which has no preceding element.

While the choice of $\mathfrak{d}_i$ should generally ensure that $j - \mathfrak{d}_i$ remains small for efficient encoding of the leftmost element, a desirable choice may vary depending on the application and sparsity pattern.
When assuming that the matrix has a banded structure, commonly obtained via (reverse) Cuthill–McKee ordering, a natural choice would be
\begin{equation}
  \label{eq:-left-offset}
  \mathfrak{d}_i = \begin{cases}
    i - k_\mathrm{left} & \text{if } k_\mathrm{left} < i, \\
    0                   & \text{otherwise},
  \end{cases}
\end{equation}
where $k_\mathrm{left}$ denotes the lower bandwidth of the matrix.
With this definition, $\mathfrak{d}_i$ can be computed directly from the row index by storing only $k_\mathrm{left}$.
In this work, we employ a slight variation of this definition (see \cref{subsec:sell-style-alignment} for details).

\subsection{Delta–Value Packing} \label{sub:packing}
While delta encoding can reduce index size, \proposed{} further decreases the memory footprint by packing each delta together with its corresponding matrix value into a single $W$-bit word.
Each word consists of a $D$-bit unsigned integer for the delta and $V$ bits for the value.
By adjusting the allocation of $D$ and $V$, \proposed{} supports a wide range of data formats for matrix values.
When required (e.g., in SpMV), the packed delta and value are efficiently unpacked on the fly.

\subsubsection{Word Structure}
A possible problem of this packing scheme is that some deltas may overflow $D$ bits, especially when $D$ is small.
Depending on the sparsity pattern and matrix size, large deltas may arise and force an increase in $D$ to prevent the overflow.
This, in turn, reduces $V$ and limits the variety of possible data representations.

\begin{figure}[t]
  \centering
  \includegraphics[width=0.65\linewidth]{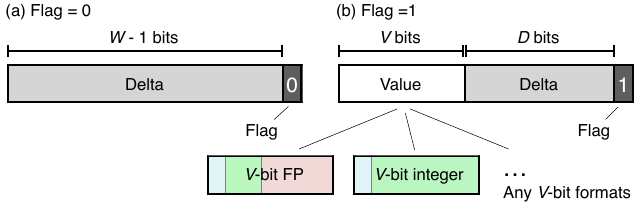}
  \caption{Structure of a word in the \proposed{} format.}
  \label{fig:packsell-word}
\end{figure}

To address this, \proposed{} introduces a flag bit at the least significant bit (\abbrev{LSB}) and employs two encoding types depending on the flag value, as illustrated in \cref{fig:packsell-word};
the bit allocation satisfies $W = V + D + 1$.
When a delta is smaller than $2^{D}$, it is stored together with its corresponding matrix value using $D$ and $V$ bits, respectively.
Specifically, the word is encoded with flag = 1 as shown in \cref{fig:pack-unpack-frame1}, which assumes $W = 32$.
The value is converted to a $V$-bit representation and stored in the upper $V$ bits, while the delta is shifted left by one bit to accommodate the flag in the LSB.
The value can be represented in any $V$-bit data format, including non-\ieee{} floating-point or integer formats.

By contrast, when a delta exceeds $2^{D}-1$, all $W - 1 (= V+D)$ bits are used to store the delta (shifted left by one bit as \lstinline[style=cppstyle-inline]{delta << 1}) with flag = 0.
As a result, the maximum representable delta becomes $2^{W-1}-1$, equivalent to that of a $W$-bit signed integer.
In this case, however, the corresponding matrix value cannot be represented.
This issue is addressed by introducing a dummy element in the sparse matrix format; see \cref{ssub:example-pack-unpack}.

This design enables a branch-free unpacking process, as shown in \cref{fig:pack-unpack-frame2}, which is especially critical for high-performance SpMV on GPUs.
First, the flag is extracted to determine the presence of a matrix value.
Second, the shift width to recover the delta is adjusted accordingly: $V (= 32-D-1)$ bits when flag = 1, and 0 bits when flag = 0.
The delta is then extracted by applying left shifting followed by right shifting, based on this shift width.
Similarly, the upper $V$ bits corresponding to the value are extracted and multiplied by the flag to drop to zero when the value is absent.
Finally, the extracted bits are reinterpreted as the target working datatype.

\begin{figure}[t]
  \subfloat[Packing\label{fig:pack-unpack-frame1}]{
    \begin{minipage}{0.95\linewidth}
\begin{lstlisting}[style=cppstyle, frame=lines]
template <typename T> uint32_t
packing(T value, uint32_t delta, uint8_t D) {
  uint32_t _value = /*
    Convert value to a V-bit representation and
    store it in the upper V bits of uint32_t. */;
  return _value | delta << 1 | 0x00000001;
}
\end{lstlisting}
    \end{minipage}
  }%
  \hfill
  \subfloat[Unpacking\label{fig:pack-unpack-frame2}]{
    \begin{minipage}{0.95\linewidth}
\begin{lstlisting}[style=cppstyle, frame=lines]
template <typename T> std::pair<T, uint32_t>
unpacking(uint32_t pack, uint8_t D) {
  uint32_t flag = pack & 0x00000001;
  uint32_t shift = (31 - D) * flag;
  uint32_t delta = (pack << shift) >> (shift + 1);
  uint32_t _value = 
    pack & ~((1u << (D + 1)) - 1) * flag;
  T value = /* Reinterpret _value as type T. */
  return {value, delta};
}
\end{lstlisting}
    \end{minipage}
  }
  \caption{Pseudocode for the packing and unpacking processes in CUDA/C++.}
  \label{fig:pack-unpack-frame}
\end{figure}

\subsubsection{Examples of Packing and Unpacking} \label{ssub:example-pack-unpack}
Although arbitrary $V$-bit data representations can be used in the packing scheme, efficient computation generally requires compatibility with hardware-supported formats.
That is, stored values should be easily convertible to such formats.
In practice, formats compatible with \ieee{} floating-point formats are particularly desirable, as they can be efficiently processed by modern arithmetic units.

We illustrate two detailed examples of packing and unpacking for $W = 32$, which are used in \cref{sec:numerical}.
First, we explain the case where $V = 16$ and FP16 is used directly.
Here, the upper 16 bits of the 32-bit word store an FP16 value.
On lines 3--5 in \cref{fig:pack-unpack-frame1}, after format conversion to FP16, the value is directly placed in the upper bits as
\begin{lstlisting}[style=cppstyle]
uint32_t _value = static_cast<uint32_t>(
  std::bit_cast<uint16_t>(static_cast<_half>(value))
) << 16;
\end{lstlisting}
Unpacking is also straightforward.
On line 8 in \cref{fig:pack-unpack-frame2}, the extracted upper 16 bits are reinterpreted as an FP16 value using the CUDA FP16 datatype \lstinline[style=cppstyle-inline]{__half}:
\begin{lstlisting}[style=cppstyle]
__half value = std::bit_cast<__half>( 
  static_cast<uint16_t>(pack >> 16));
\end{lstlisting}

Next, we consider using an E8M$Y$ floating-point format for $V$ bits, which consists of one sign bit, 8 exponent bits, and $Y(= 22-D)$ mantissa bits.
E8M$Y$ is compatible with FP32 and preserves key properties such as the exponent bias and implicit leading 1.
In this case, to convert FP32 values into E8M$Y$, we use a combination of rounding to the nearest integer and bitwise truncation on lines 3--5 in \cref{fig:pack-unpack-frame1}:
\begin{lstlisting}[style=cppstyle]
int exp;
frexpf(value, &exp); // assuming a float value
float scale = ldexpf(1.0, exp - 24 + D + 1);
float __value = std::round(value / scale) * scale;
uint32_t _value = 
  (std::bit_cast<uint32_t>(__value) >> (D + 1)) 
  << (D + 1);
\end{lstlisting}
Unpacking is simpler than the FP16 case, as the extracted $V$ bits can be directly interpreted as an FP32 value.
Line 8 in \cref{fig:pack-unpack-frame2} corresponds to
\begin{lstlisting}[style=cppstyle]
float value = std::bit_cast<float>(_value);
\end{lstlisting}

\subsection{SELL-style Alignment} \label{subsec:sell-style-alignment}
Our delta–value packing scheme allows \proposed{} to handle large deltas of up to $2^{W-1}-1$ with flag = 0.
However, in this case, the corresponding matrix value cannot be stored together with the delta.
To address this, when the delta between two consecutive nonzero elements exceeds the range of the $D$-bit unsigned integer, \proposed{} inserts a dummy element at the same column index as the target element.
Since the dummy element has no effective value, it can be encoded with flag = 0.
The actual nonzero element then has a delta of 0 relative to the dummy element and can be encoded with flag = 1.
This design allows $D$ to be reduced, potentially to as small as 1 bit, at the cost of additional dummy elements.
That is, it increases the choice of $V$ for representing matrix values, thereby improving the variation of data representations.

After inserting dummy elements for all large deltas, \proposed{} applies SELL-style alignment with slice size $C$ to both the original and dummy elements without distinction.
Because deltas and values are packed into $W$-bit words, the \texttt{val} and \texttt{col} arrays in SELL are reduced to a single array, \texttt{pack}, in \proposed{}, along with \texttt{offset} for slice boundaries.
For this reason, the word size $W$ should be selected considering the memory access efficiency for \texttt{pack} (e.g., 64, 32).

\begin{figure}[t]
  \centering
  \includegraphics[width=0.6\linewidth]{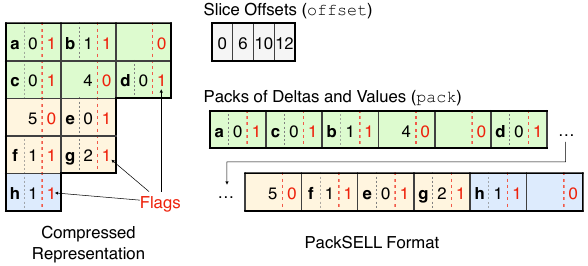}
  \caption{Example of the \proposed{} format with slice size $C = 2$ and bit-width for deltas $D = 2$. For each row, $\mathfrak{d}_i = 0$ for simplicity.}
  \label{fig:packsell-format}
\end{figure}

\Cref{fig:packsell-format} illustrates an example of \proposed{} with $D = 2$ for the same sample matrix as shown in \cref{fig:sell}, where $\mathfrak{d}_i$ is assumed to be 0 for simplicity.
In this example, only deltas in the range of 0 to 3 are directly representable.
For example, since the delta between entries $c$ and $d$ is 4, a dummy element is inserted immediately before $d$.
In addition, padding elements introduced by SELL-style alignment are treated as zero-valued entries and encoded with flag = 0 and delta = 0, which remains consistent with the delta–value packing scheme.

Similar to SELL-$C$-$\sigma$, \proposed{} supports row permutation within blocks of $\sigma$ rows to reduce padding by balancing the number of stored elements across rows within each slice.
Most aspects of the row permutation follow those of SELL-$C$-$\sigma$, including the memory layout and the procedure for restoring the original ordering.
However, since \proposed{} introduces dummy elements for large deltas and applies SELL-style padding after their insertion, the row permutation also accounts for these inserted dummy elements.

When using implicit permutation, the offset for the leftmost nonzero elements, $\mathfrak{d}_i$, should be identical within each block of $\sigma$ rows.
Otherwise, to obtain $\mathfrak{d}_i$ after the permutation, all $\mathfrak{d}_i$ values or the inverse permutation must be stored even when using the definition \cref{eq:-left-offset}, leading to additional memory accesses.
To avoid this, we define $\mathfrak{d}_i$ uniformly within each block as:
\begin{equation}
  \label{eq:left-offset}
  \mathfrak{d}_i = \begin{cases}
    \left\lfloor i / \sigma \right\rfloor \cdot \sigma - k_\mathrm{left} & \text{if } k_\mathrm{left} < \left\lfloor i / \sigma \right\rfloor \cdot \sigma, \\
    0                                                                    & \text{otherwise}.
  \end{cases}
\end{equation}

\subsection{\proposed{}-based SpMV} \label{sub:packsell-spmv}
SpMV in \proposed{} can be implemented similarly to the SELL-based implementation in \cref{sec:sell}, with the addition of on-the-fly unpacking.
While further hardware-aware optimizations may be possible, a straightforward implementation is shown below and is used in the numerical evaluations:
\begin{algorithmic}[1]
  \For{$i = 0,\ldots,n-1$}
  \State $c := \mathfrak{d}_{i}$ \Comment{Use \cref{eq:left-offset}}
  \State $i' := \lfloor i / \sigma \rfloor \cdot \sigma + \texttt{perm}_i$
  \State $k := \lfloor i / C \rfloor$
  \State $l := i \bmod C$
  \State $s := \texttt{offset}_k$
  \State $w := (\texttt{offset}_{k + 1} - s) / C$
  \State $t := 0$
  \For{$j=0,\ldots,w-1$}
  \State $p := s+j \cdot C+l$
  \State Unpack $\texttt{pack}_p$ to obtain $(v, d)$
  \State $c := c + d$
  \State $t := t + v \cdot x_{c}$
  \EndFor
  \State $y_{i'} = t$
  \EndFor
\end{algorithmic}
Efficient unpacking is critical for high performance.
In this study, we implement it as explained in \cref{sub:packing}.

\section{Numerical Experiments} \label{sec:numerical}
We evaluate \proposed{} from two perspectives: the performance of standalone SpMV kernels (\cref{subsec:eval-spmv}) and their performance when integrated into Krylov subspace methods for solving sparse linear systems (\cref{subsec:eval-solver}).
Although \proposed{} supports multiple word sizes $W$ (e.g., 16, 32, and 64), we focus on $W = 32$.
This setting is comparable to common low-precision SpMV implementations that use FP32 or FP16 values with 32-bit (signed) integer indices.

All experiments were conducted on a GPU node of the Gardenia supercomputer at Kyoto University.
Although each node has four NVIDIA A100 (80GB SXM) GPUs, we used a single GPU.
The A100 GPU provides a memory bandwidth of 2,039 GB/s, and the installed driver version was 580.105.08.
The evaluated kernels and solvers were developed in CUDA and C++ using the NVIDIA CUDA Compiler (version 13.0.2) and the GNU C++ Compiler (version 13.3.0), with the options \texttt{-O3}, \texttt{-sm\_80}, \texttt{-std=c++20}, and \texttt{-Xcompiler="-O3 -fopenmp -std=c++20"}.
The C++20 standard was required to support advanced bit manipulation.
We also used the cuSPARSE library from the same CUDA Toolkit.

\subsection{Evaluation via Standalone SpMV Kernels}
\label{subsec:eval-spmv}
This subsection evaluates standalone SpMV kernels.
To reflect typical use cases of GPUs for low-precision SpMV, we used relatively large matrices from the SuiteSparse Matrix Collection~\citep{davis2011UniversityFlorida}.
Specifically, we selected most large-scale real (and integer) matrices with at least 65,536 (= $2^{16}$) columns.
Since precision is a key focus of this study, we excluded binary matrices having only 0 and 1 values.
Very small matrices were also excluded, as they are inefficient on modern GPUs and their column indices can be represented with 16-bit or 8-bit integers.
Among the 438 matrices satisfying these conditions, three matrices (\mname{MOLIERE_2016}, \mname{GAP-urand}, and \mname{GAP-kron}) were excluded due to memory limitations.
Consequently, 435 matrices were used in the evaluation.

All performance results in this subsection are averages of 10,000 runs after 100 warm-up runs.
In addition, FLOPS is measured assuming two floating-point operations per nonzero element, excluding padding and dummy elements.

\subsubsection{FP16 SpMV}
\label{ssub:eval-spmv-fp16}
First, we examine \proposed{} for FP16 SpMV, where the input and output vectors are also stored in FP16.
Based on the algorithm in \cref{sub:packsell-spmv}, we developed an FP16 SpMV kernel in \proposed{} with $W=32$ and $D=15$;
FP16 values were directly embedded into the remaining $V(= 16)$ bits, as explained in~\cref{ssub:example-pack-unpack}.
The kernel was parallelized such that each thread processes one row, with 256 CUDA threads per block.

This kernel was compared with five FP16 SpMV kernels: DASP~\citep{lu2023DASPSpecific} and four NVIDIA cuSPARSE kernels\footnote{Although cuSPARSE provides an interface for computing $y = \alpha Ax + \beta y$ for scalars $\alpha$ and $\beta$, we observed better performance when setting $\beta = 0$.} for COO, CSR, SELL, and BSR formats, hereafter denoted as cuCOO, cuCSR, cuSELL, and cuBSR, respectively.
All cuSPARSE kernels were used with \texttt{cusparseSpMV\_preprocess()} for potential optimization.
BSR used either 2$\times$2 or 4$\times$4 blocks, whichever performed better.
For \proposed{} and cuSELL, we fixed $C=32$ (warp size) and $\sigma=256$, and explicitly reordered matrix rows, since cuSELL, the closest counterpart to \proposed{}, does not support implicit permutation.
Reordering the output vector to restore the original ordering would incur additional memory accesses and penalize cuSELL.

\begin{figure}[tbp]
  \centering
  \includegraphics[width=0.65\linewidth]{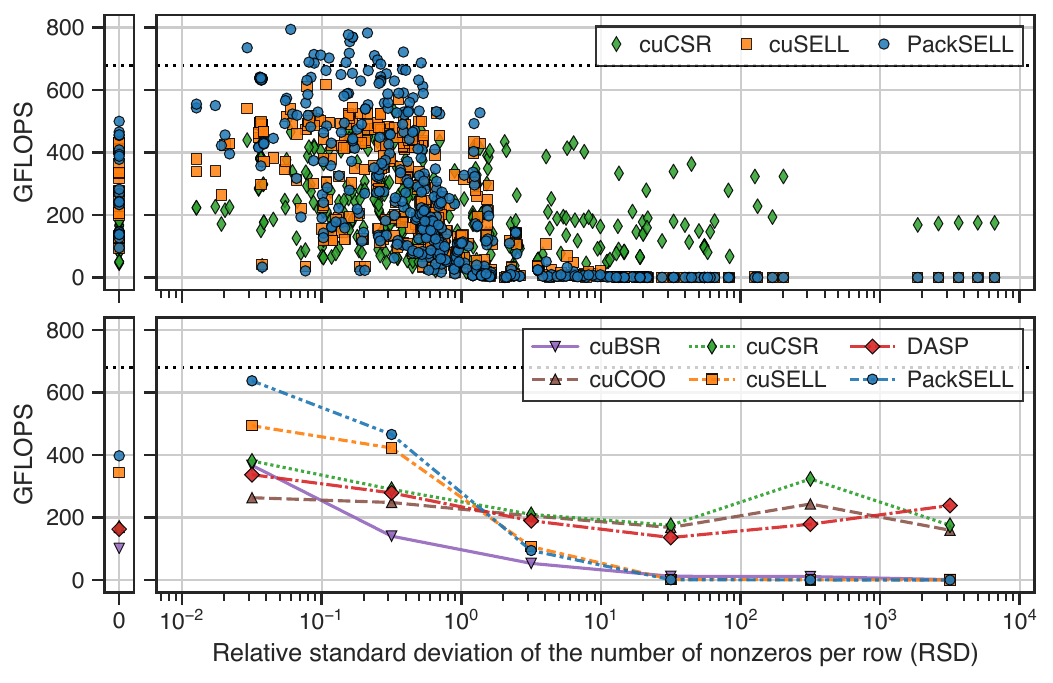}
  \caption{Achieved FLOPS for six SpMV kernels. Dotted horizontal lines indicate the upper bound based only on nonzero elements (FP16 values and 32-bit column indices).}
  \label{fig:fp16-flops}
\end{figure}

\begin{figure}[tbp]
  \centering
  \includegraphics[width=0.65\linewidth]{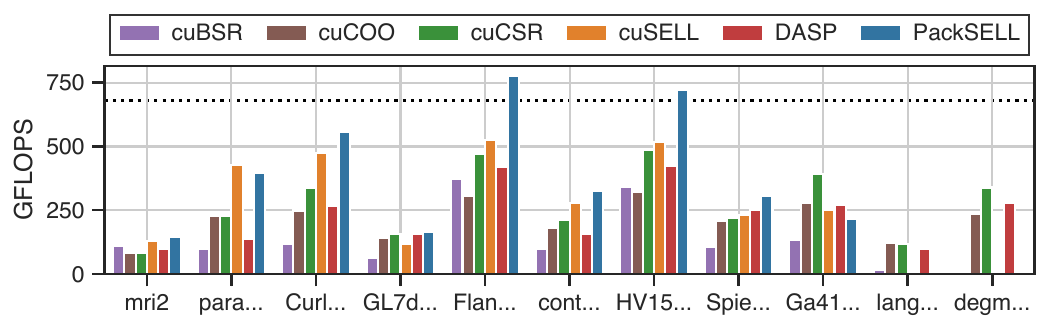}
  \caption{Detailed results for matrices listed in \cref{tab:fp16-mat}.}
  \label{fig:fp16-select}
\end{figure}

\Cref{fig:fp16-flops} summarizes the achieved FLOPS of each kernel.
The horizontal axis shows the relative standard deviation of the number of nonzero elements per row (\abbrev{RSD}), where smaller values indicate a more uniform distribution of nonzero elements across all rows.
The upper plot shows results for individual matrices, and the lower plot reports the 75th percentile for seven bins (RSD = 0 or in $[10^{e}, 10^{e+1})$ for $e \in [-2, 3]$) to highlight representative performance.
To further illustrate the relationship between matrix characteristics and performance, \cref{fig:fp16-select} presents detailed results for 11 matrices listed in \cref{tab:fp16-mat}, selected to represent diverse sparsity structures.

\begin{table}[t]
    \caption{Selected Matrices}
    \label{tab:fp16-mat}
    \centering
    \footnotesize
    \setlength{\tabcolsep}{3pt}
    \begin{threeparttable}
        \begin{tabular}{lcrrr}
            \hline
            \makecell[l]{Matrix                \\ \,\,/ \scriptsize{Application}} & Sparsity & \makecell[r]{$n$ $\times$ $m$} & $n_{nz}$\tnote{\dag} & RSD \\
            \hline
            \makecell[l]{\mname{mri2}          \\ \,\,/ \scriptsize{Computer graphics}} & \includegraphics[width=3em,height=3em,keepaspectratio,valign=c]{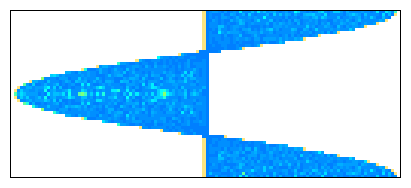}        & \makecell[r]{\num{63240} \\$\times$\num{147456}}  & \num{569160}  & 0      \\
            \makecell[l]{\mname{parabolic_fem} \\ \,\,/ \scriptsize{CFD}} & \includegraphics[width=3em,height=3em,keepaspectratio,valign=c]{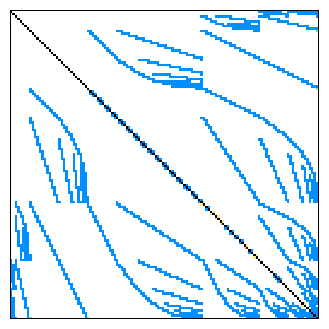}             & \makecell[r]{\num{525825} \\$\times$\num{525825}} & \num{3674625} & 0.0218 \\
            \makecell[l]{\mname{CurlCurl_4}    \\ \,\,/ \scriptsize{Model reduction}} & \includegraphics[width=3em,height=3em,keepaspectratio,valign=c]{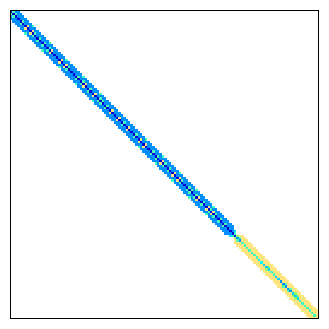}    & \makecell[r]{\num{2380515} \\$\times$\num{2380515}} & \num{26515867} & 0.0768 \\
            \makecell[l]{\mname{GL7d17}        \\ \,\,/ \scriptsize{Differential matrix}} & \includegraphics[width=3em,height=3em,keepaspectratio,valign=c]{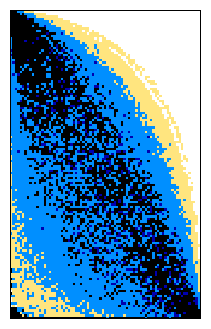}  & \makecell[r]{\num{1548650} \\$\times$\num{955128}} & \num{25978098} & 0.1180 \\
            \makecell[l]{\mname{Flan_1565}     \\ \,\,/ \scriptsize{Structural analysis}} & \includegraphics[width=3em,height=3em,keepaspectratio,valign=c]{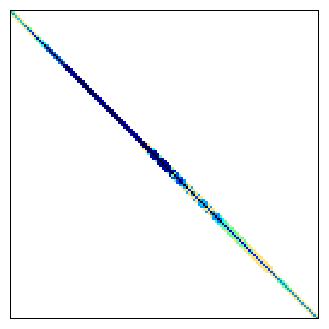}  & \makecell[r]{\num{1564794} \\$\times$\num{1564794}} & \num{114165372} & 0.1550 \\
            \makecell[l]{\mname{cont11_l}      \\ \,\,/ \scriptsize{Linear programming}} & \includegraphics[width=3em,height=3em,keepaspectratio,valign=c]{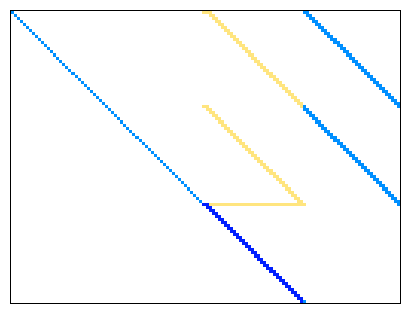}   & \makecell[r]{\num{1468599} \\$\times$\num{1961394}} & \num{5382999} & 0.2571 \\
            \makecell[l]{\mname{HV15R}         \\ \,\,/ \scriptsize{CFD}} & \includegraphics[width=3em,height=3em,keepaspectratio,valign=c]{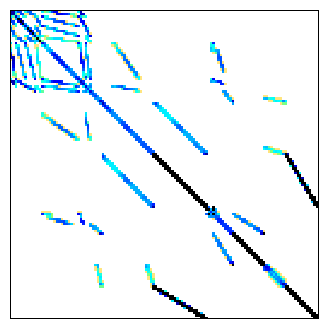}                     & \makecell[r]{\num{2017169} \\$\times$\num{2017169}} & \num{283073458} & 0.3845 \\
            \makecell[l]{\mname{Spielman_k600} \\ \,\,/ \scriptsize{Chimera Laplacian}} & \includegraphics[width=3em,height=3em,keepaspectratio,valign=c]{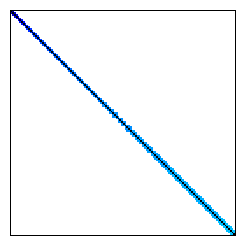}  & \makecell[r]{\num{72180402} \\$\times$\num{72180402}} & \num{216902404} & 0.6655 \\
            \makecell[l]{\mname{Ga41As41H72}   \\ \,\,/ \scriptsize{Quantum chemistry}} & \includegraphics[width=3em,height=3em,keepaspectratio,valign=c]{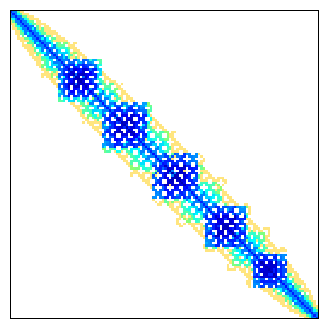} & \makecell[r]{\num{268096} \\$\times$\num{268096}} & \num{18488476} & 1.5282 \\
            \makecell[l]{\mname{language}      \\ \,\,/ \scriptsize{Weighted graph}} & \includegraphics[width=3em,height=3em,keepaspectratio,valign=c]{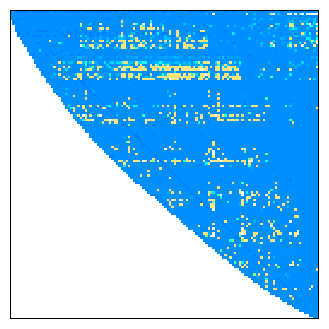}       & \makecell[r]{\num{399130} \\$\times$\num{399130}} & \num{1216334} & 6.7958 \\
            \makecell[l]{\mname{degme}         \\ \,\,/ \scriptsize{Linear programming}} & \includegraphics[width=3em,height=3em,keepaspectratio,valign=c]{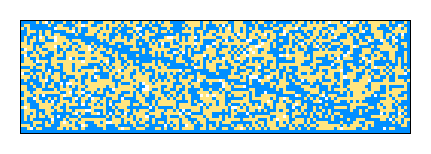}       & \makecell[r]{\num{185501} \\$\times$\num{659415}} & \num{8127528} & 33.0696 \\
            \hline
        \end{tabular}
        \begin{tablenotes}
            \item[\dag] $n_{nz}$ denotes the number of nonzero elements.
        \end{tablenotes}
    \end{threeparttable}
\end{table}

These results first reconfirm the strengths and limitations of SELL-based formats.
In general, SELL is suitable for matrices with regular sparsity patterns (i.e., small RSD).
For such matrices, cuSELL approached the roofline peak more closely than other kernels, such as cuCSR and cuCOO.
By contrast, for matrices with high RSD values, such as \mname{Ga41As41H72}, \mname{language}, and \mname{degme}, SELL incurs substantial padding and underperforms CSR and COO, which require no padding.
Accordingly, cuSELL and \proposed{} underperformed cuCSR and cuCOO in these cases, which indicates a limitation of SELL-based approaches.
Achieving higher performance for such matrices remains future work.
Nevertheless, among the 435 matrices tested, cuSELL outperformed cuCSR for 247 matrices; this number increases to 292 when including cases where cuSELL achieved at least 90\% of the performance of cuCSR.
These results indicate that improving upon cuSELL is practically meaningful.
We refer to these 292 matrices as \emph{\suitable{} matrices} and focus on them in the following.

\begin{figure}[tbp]
  \centering
  \includegraphics[width=0.65\linewidth]{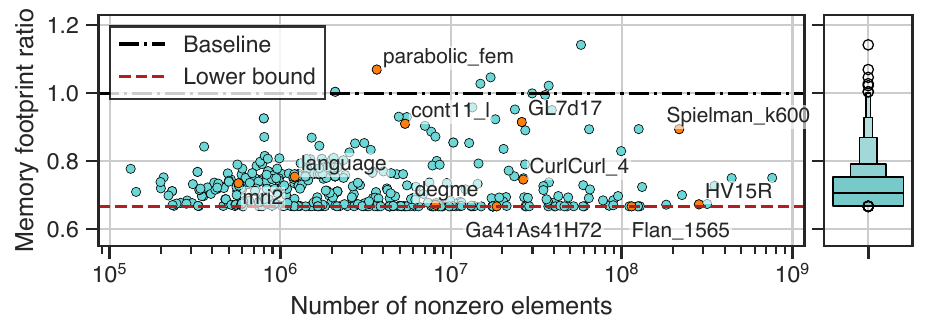}
  \caption[Memory footprint ratio of \proposed{} to SELL.]{Memory footprint ratio of \proposed{} to SELL, shown as scatter and letter-value plots. Orange markers indicate matrices listed in \cref{tab:fp16-mat}.}
  \label{fig:fp16-storage}
\end{figure}

For many \suitable{} matrices, \proposed{} outperformed cuSELL and exceeded the upper bound of cuCSR and cuSELL by reducing the memory footprint.
In particular, \proposed{} achieved larger gains for matrices with small RSD and many nonzero elements, such as \mname{CurlCurl_4}, \mname{Flan_1565}, and \mname{HV15R}.
\Cref{fig:fp16-storage} further explains these results showing the memory usage of \proposed{} relative to SELL.
In most cases, \proposed{} reduced the memory footprint compared to SELL via delta–value packing, with few dummy elements.
In particular, when many nonzero elements are densely clustered, as in the three examples above, $D$-bit (15-bit) unsigned integers can represent most deltas, enabling compression rates close to the lower bound of 0.75 (= 32 bits / 48 bits).

By contrast, when nonzero elements were widely scattered, the number of dummy elements increased, and the performance gain decreased.
In such cases, except for small matrices like \mname{mri2}, where $D$ bits still covered most deltas, \proposed{} required a similar amount of storage to cuSELL (e.g., \mname{cont11_l} and \mname{GL7d17}), and even underperformed cuSELL due to the increased storage (e.g., \mname{parabolic_fem}).
These results suggest that matrix reordering to improve the locality of nonzero elements is promising for further improvements of \proposed{}, which we leave for future work.

\begin{figure}[tbp]
  \centering
  \includegraphics[width=0.65\linewidth]{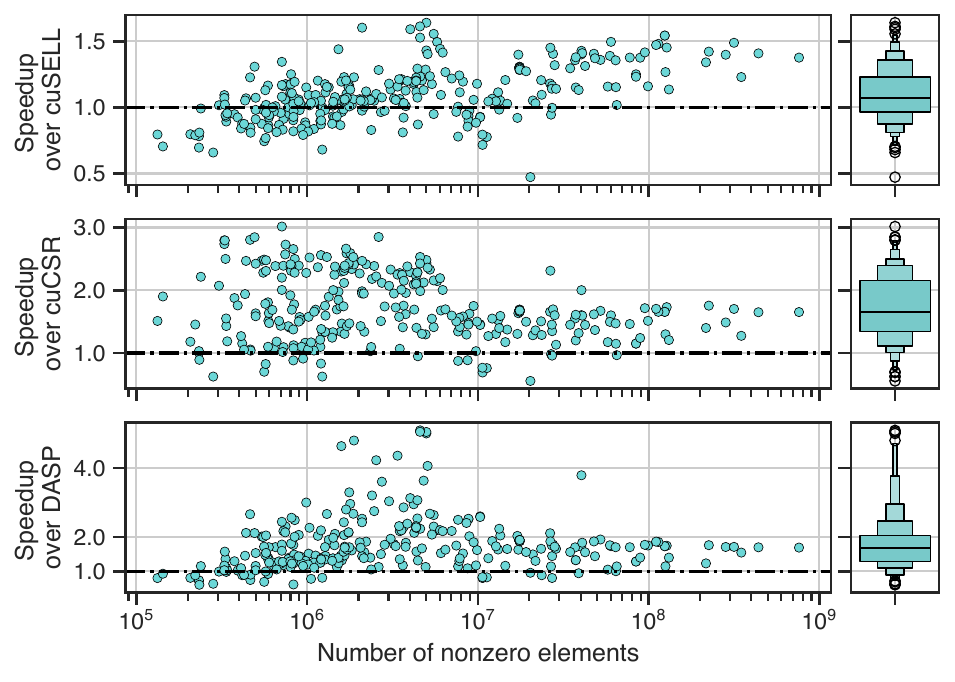}
  \caption{Speedups of \proposed{} over cuSELL, cuCSR, and DASP for the \suitable{} matrices.}
  \label{fig:fp16-speedup}
\end{figure}

\cref{fig:fp16-speedup} shows the speedups of \proposed{} for the \suitable{} matrices.
\proposed{} outperformed cuSELL on 201 of 292 matrices.
For matrices with many nonzero elements, it achieved consistent speedups of around $1.5\times$, matching the ideal gain expected from the reduced data size.
In some cases, speedups reached up to $1.63\times$, likely due to improved memory access patterns from delta–value packing, which may enhance cache efficiency by using a single array for values and deltas.
On smaller matrices where even cuSELL struggled to fully utilize memory bandwidth, \proposed{} underperformed cuSELL;
addressing these cases with hardware-oriented optimizations, such as those to effectively hide memory latency, remains future work.
Compared with cuCSR and DASP, \proposed{} outperformed on 274 and 265 matrices, respectively, with maximum speedups of 3.01$\times$ and 5.09$\times$.
For reference, the maximum speedups of \proposed{} against cuCOO and cuBSR were 2.76$\times$ and 7.57$\times$, respectively.

\subsubsection{SpMV Using Non-IEEE Formats} \label{ssub:spmv_for_various_formats}
We next assess the advantage of \proposed{} in supporting various data formats.
We tested \proposed{}-based SpMV while varying $D$ from 1 to 12.
The implementation in each case was mostly the same as that in \cref{ssub:eval-spmv-fp16}, expect that matrix values were represented in E8M$Y$ with $Y = 22 - D$.
We compared these kernels against FP32 cuSELL; in both cases, the input and output vectors were stored in FP32.
In addition to performance, we evaluated accuracy using the backward error, defined as
\begin{equation}
  \frac{\|y - Ax\|}{\|A\|\|x\|},
\end{equation}
where the infinity norm was used.
Although acceptable values of the backward error depend on the application, comparing backward errors across different precision settings provides a relative measure of practicality.
For more practical evaluations using iterative linear solvers, see the next subsection.

For a focused evaluation, we restricted our analysis to the 292 SELL-suitable matrices.
To mitigate overflow and underflow, we applied diagonal scaling $G^{-1}A$, where $G = \mathrm{diag}(g_1,\ldots,g_n)$ with $g_i = \Sigma_j |a_{i,j}|$.

\begin{figure*}[tbp]
  \centering
  \includegraphics[width=0.9\linewidth]{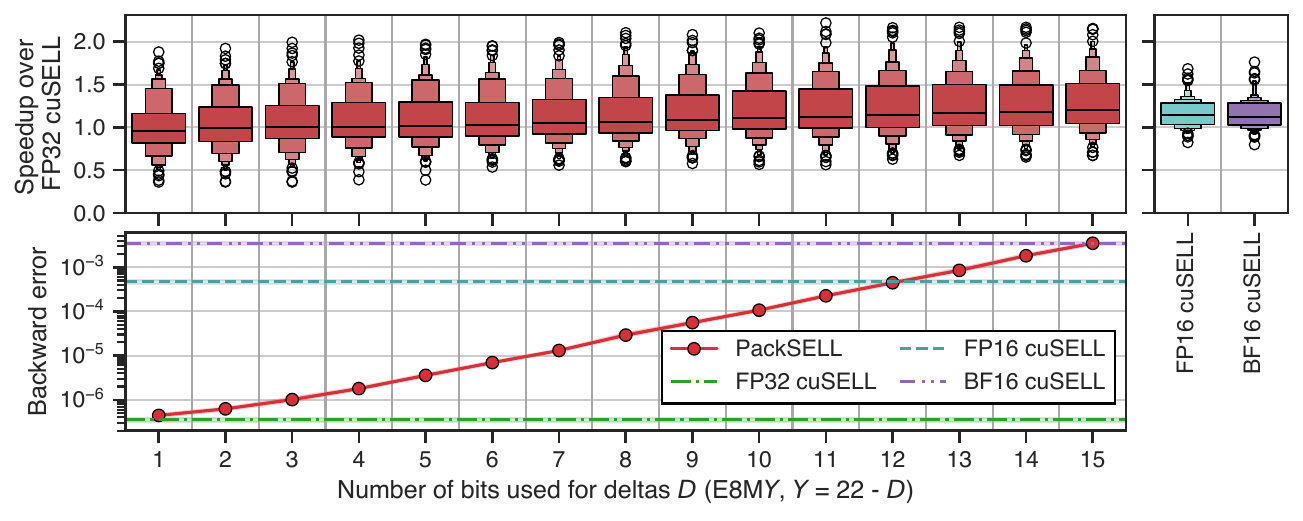}
  \caption{Achieved performance and backward error of cuSELL and \proposed{}-based SpMV using E8M$Y$. The reported backward error is the average across all matrices, excluding the result of FP16 cuSELL for \mname{t3dh_e}, which is omitted due to an outlier caused by underflow.}
  \label{fig:various}
\end{figure*}

The performance of \proposed{} relative to FP32 cuSELL is summarized in \Cref{fig:various} using letter-value plots, along with the average backward error over all matrices.
For reference, the figure also includes results for cuSELL in FP16 and bfloat16 (\abbrev{BF16}) with FP16 or BF16 inputs and FP32 outputs.

As expected, increasing $D$ (and thus decreasing $Y$) improved performance by reducing the number of dummy elements for large deltas, at the cost of lower precision.
Although the backward error gradually increased as the number of mantissa bits $Y$ decreased, the degradation was moderate, especially when only a few bits were truncated;
in most cases, the increase remained within one order of magnitude.

In contrast to the comparable backward error, \proposed{} with small $D$ outperformed cuSELL in many cases, e.g., 129 out of 292 matrices for $D=1$ (E8M21) and 145 matrices for $D=2$ (E8M20).
It also delivered notable speedups, approaching 2.0$\times$ in some instances.
Furthermore, under these settings, \proposed{} even outperformed FP16/BF16 cuSELL, while achieving much lower backward errors than both.
As observed in \cref{ssub:eval-spmv-fp16}, the performance of \proposed{} is affected by the locality of nonzero elements.
These results suggest that, for matrices suited to \proposed{} with high locality of nonzero elements, even a few $D$ bits suffice to represent deltas and yield both higher accuracy and improved performance.

Another notable case is $D=12$ (E8M10).
Despite allocating 8 bits to the exponent (and using an FP32 input vector), \proposed{} outperformed FP16 cuSELL for 175 matrices.
For matrices with a wide dynamic range, a larger exponent part can be beneficial.
Indeed, FP16 cuSELL suffered severe underflow for the \mname{t3dh_e} matrix, with a backward error of $8.6\times10^{-1}$, even with diagonal scaling (this case is excluded from \cref{fig:various} as an outlier).
In those cases, BF16 might act as an alternative; however, the performance of BF16 cuSELL is similar to that of FP16 cuSELL, and its 7-bit mantissa can be insufficient in certain applications.
These results further emphasize the key benefit of \proposed{}: the ability to flexibly adjust exponent and mantissa bits according to the characteristics of matrices and applications.

\subsection{Evaluation via Krylov Solvers} \label{subsec:eval-solver}
This subsection evaluates the practical performance of \proposed{}-based SpMV kernels in Krylov subspace methods for solving sparse linear systems $Ax = b$.
Specifically, we consider two mixed-precision solvers: the state-of-the-art FP16-enabled solver F3R~\citep{suzuki2025NestedKrylov} and an inner-outer conjugate gradient (IO-CG) method.

The experiments used 30 square matrices listed in \cref{tab:matrices}.
Of these, 22 belong to the previously described set of 292 \suitable{} matrices, while the remaining 8 are derived from the HPCG~\citep{dongarra2016Highperformanceconjugategradient} and HPGMxP~\citep{yamazaki2022HighPerformanceGMRES} benchmarks.
These matrices are denoted as \mname{HPCG_x_y_z} and \mname{HPGMP_x_y_z}, where the number of rows $n$ is given by $2^{\texttt{x}+\texttt{y}+\texttt{z}}$.
For HPGMxP, the parameter to control the asymmetry was set to 0.5.
While F3R experiments used all 30 matrices, IO-CG experiments used only the 15 symmetric positive definite (SPD) matrices since CG is designed for SPD systems.
For all matrices, diagonal scaling of the form $\bar{G}^{-1}A\bar{G}^{-1}$ was applied, where $\bar{G} = \mathrm{diag}(\bar{g}_1,\ldots,\bar{g}_n)$ with $\bar{g}_i = \sqrt{|a_{i,i}|}$.

\begin{table}[t]
    \caption{Test Matrices: 15 Symmetric Positive Definite Matrices (Top) and 15 Nonsymmetric Matrices (Bottom)}
    \label{tab:matrices}
    \centering
    \footnotesize
    \setlength{\tabcolsep}{5pt}
    \begin{threeparttable}
        \begin{tabular}{lrrrr}
            \hline
            Matrix                & $n$            & $n_{nz}$        & $n_{nz}$/$n$ & $\kappa_2$\tnote{\ddag} \\
            \hline
            \mname{Bump_2911}     & \num{2911419}  & \num{127729899} & 43.87        & 1.5e+06                 \\
            \mname{Emilia_923}    & \num{923136}   & \num{40373538}  & 43.74        & 1.8e+06                 \\
            \mname{G3_circuit}    & \num{1585478}  & \num{7660826}   & 4.83         & 2.4e+05                 \\
            \mname{Queen_4147}    & \num{4147110}  & \num{316548962} & 76.33        & 9.4e+06                 \\
            \mname{Serena}        & \num{1391349}  & \num{64131971}  & 46.09        & 2.7e+04                 \\
            \mname{apache2}       & \num{715176}   & \num{4817870}   & 6.74         & 1.3e+06                 \\
            \mname{audikw_1}      & \num{943695}   & \num{77651847}  & 82.28        & 1.4e+07                 \\
            \mname{ecology2}      & \num{999999}   & \num{4995991}   & 5.00         & 6.3e+07                 \\
            \mname{ldoor}         & \num{952203}   & \num{42493817}  & 44.63        & 4.1e+06                 \\
            \mname{thermal2}      & \num{1228045}  & \num{8580313}   & 6.99         & 4.5e+06                 \\
            \mname{tmt_sym}       & \num{726713}   & \num{5080961}   & 6.99         & 2.5e+08                 \\
            \mname{HPCG_7_7_7}    & \num{2097152}  & \num{55742968}  & 26.58        & 2.3e+03                 \\
            \mname{HPCG_8_7_7}    & \num{4194304}  & \num{111777784} & 26.65        & 3.0e+03                 \\
            \mname{HPCG_8_8_7}    & \num{8388608}  & \num{224140792} & 26.72        & 4.5e+03                 \\
            \mname{HPCG_8_8_8}    & \num{16777216} & \num{449455096} & 26.79        & 8.9e+03                 \\
            \hline
            \mname{Freescale1}    & \num{3428755}  & \num{17052626}  & 4.97         & 3.8e+07                 \\
            \mname{Transport}     & \num{1602111}  & \num{23487281}  & 14.66        & 4.0e+05                 \\
            \mname{atmosmodd}     & \num{1270432}  & \num{8814880}   & 6.94         & 5.2e+03                 \\
            \mname{atmosmodl}     & \num{1489752}  & \num{10319760}  & 6.93         & 1.1e+03                 \\
            \mname{rajat31}       & \num{4690002}  & \num{20316253}  & 4.33         & 7.9e+03                 \\
            \mname{ss}            & \num{1652680}  & \num{34753577}  & 21.03        & 5.0e+04                 \\
            \mname{stokes}        & \num{11449533} & \num{349321980} & 30.51        & 5.1e+06                 \\
            \mname{t2em}          & \num{921632}   & \num{4590832}   & 4.98         & 2.3e+05                 \\
            \mname{tmt_unsym}     & \num{917825}   & \num{4584801}   & 5.00         & 1.5e+09                 \\
            \mname{vas_stokes_1M} & \num{1090664}  & \num{34767207}  & 31.88        & 4.3e+07                 \\
            \mname{vas_stokes_2M} & \num{2146677}  & \num{65129037}  & 30.34        & 6.3e+07                 \\
            \mname{HPGMP_7_7_7}   & \num{2097152}  & \num{55742968}  & 26.58        & 1.6e+03                 \\
            \mname{HPGMP_8_7_7}   & \num{4194304}  & \num{111777784} & 26.65        & 1.9e+03                 \\
            \mname{HPGMP_8_8_7}   & \num{8388608}  & \num{224140792} & 26.72        & 2.2e+03                 \\
            \mname{HPGMP_8_8_8}   & \num{16777216} & \num{449455096} & 26.79        & 4.2e+03                 \\
            \hline
        \end{tabular}
        \begin{tablenotes}
            \item[\ddag] $\kappa_2$ denotes an estimated 2-norm condition number of $\bar{G}^{-1}A\bar{G}^{-1}$.
        \end{tablenotes}
    \end{threeparttable}
\end{table}

All numerical results below are averages over five runs.
In each run, the right-hand side vector $b$ was a random vector with elements uniformly distributed in the range [0, 1), and the initial guess was the zero vector.
Convergence was determined using the following criterion, where $\tilde{x}$ denotes an approximate solution:
\begin{equation}
  \|b - A\tilde{x}\|_2 / \|b\|_2 < 10^{-9}.
\end{equation}

\subsubsection{FP16-Enabled F3R Solver}
F3R is a nested Krylov subspace method that hierarchically combines Krylov solvers as preconditioners for other Krylov iterations.
It consists of four layers: three flexible GMRES (\abbrev{FGMRES}) layers and an innermost preconditioned Richardson layer.
Within this hierarchy, the innermost Richardson and its outer FGMRES employ FP16 SpMV.
Since FP16 SpMV accounts for over 85\% of all SpMV operations under the default parameter settings, implementing these kernels with \proposed{}, as described in \cref{ssub:eval-spmv-fp16}, provides an evaluation of practical performance.

We implemented F3R in three ways.
The first is an FP64-only version (FP64-F3R), and the second is a mixed-precision variant that uses FP16 SpMV in the two inner layers mentioned above (FP16-F3R).
For their SpMV operations, we used our SELL-based kernels following the algorithm in \cref{sec:sell}, with $C = 32$ and $\sigma = 256$.
This is because mixed-precision F3R requires combinations of vector and matrix value types unsupported by cuSPARSE.
Additionally, we adopted the implicit $\sigma$ permutation to preserve the original ordering, which is also not available in cuSPARSE.
The third implementation has the same structure as FP16-F3R but uses the FP16 SpMV kernel in \proposed{} ($V=16$, $D=15$) described in \cref{ssub:eval-spmv-fp16} (\proposed{}-F3R).
This implementation also employs implicit permutation with $\sigma = 256$.
In all solvers, all other parameters followed the default settings;
for example, the innermost Richardson used the approximate inverse preconditioner SD-AINV~\citep{suzuki2022NewAINV};
see~\citep{suzuki2025NestedKrylov,kengo_suzuki_2025_16882405} for details.

\begin{figure}[tbp]
  \centering
  \includegraphics[width=0.4\linewidth]{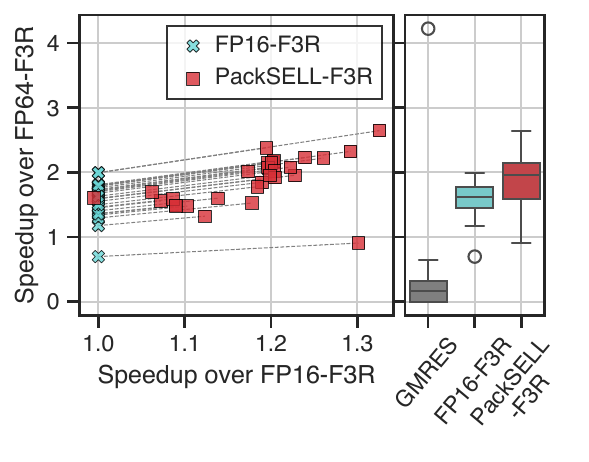}
  \caption{Performance of different F3R implementations.}
  \label{fig:f3r}
\end{figure}

The performance of these solvers is presented in \cref{fig:f3r}.
The left plot shows the speedups of \proposed{}-F3R over FP16-F3R and FP64-F3R, with dotted lines connecting results for each problem.
The right plot summarizes the speedup over FP64-F3R, with reference results for an FP64 GMRES solver combined with SD-AINV and restarted every 100 iterations.
Except for one case (\mname{rajat31}), F3R outperformed GMRES, and \proposed{} further improved performance.
Since FP16 values are directly embedded in \proposed{}, FP16-F3R and \proposed{}-F3R exhibit identical convergence.
Thus, the performance benefits in SpMV translate into gains in the overall performance.
For all problems except \mname{Freescale1}, \proposed{}-F3R outperformed FP16-F3R by 1.06--1.32$\times$.
These improvements resulted in speedups of up to 2.64$\times$ over FP64-F3R and up to 28.14$\times$ over GMRES, indicating that \proposed{} is effective in this practical scenario.

\subsubsection{Inner-Outer CG Solver}
We demonstrate that SpMV with custom data formats enabled by \proposed{} increases the flexibility of the design of mixed-precision solvers.
To illustrate this, we consider IO-CG, a variant of the mixed-precision solver presented in~\citep{buttari2008UsingMixed}.
In IO-CG, $m_\mathrm{in}$ iterations of preconditioned CG (\abbrev{PCG}) serve as a preconditioner for the outer flexible CG (\abbrev{FCG})~\citep{notay2000FlexibleConjugate}.
While this scheme is relatively uncommon, it can serve as a good example of how \proposed{} can increase the flexibility of solver design.

We developed four IO-CG variants: one FP64-only solver and three mixed-precision solvers.
In all mixed-precision variants, the outer FCG uses FP64, while the inner PCG primarily employs FP32.
These three variants differ in the SpMV kernel used for the coefficient matrix $A$: FP32 SELL, FP16 SELL, and \proposed{}, in which the values are encoded in E8M$Y$ ($Y = 22 - D$) with $D$ ranging from 1 to 12.
We refer to these solvers as FP64-, FP32-, FP16-, and E8M$Y$-IO-CG, respectively.
SpMV (and preconditioning) dominate the cost of PCG, and the cost of index access is also non-negligible when storing $A$ explicitly.
Thus, based primarily on SpMV performance, the expected ideal speedups over FP64 PCG for this approach and similar previous methods are roughly 1.5$\times$ and 2.0$\times$ when using FP32 and FP16 for SpMV, respectively.

\begin{figure}[tb]
  \centering
  \includegraphics[width=0.55\linewidth]{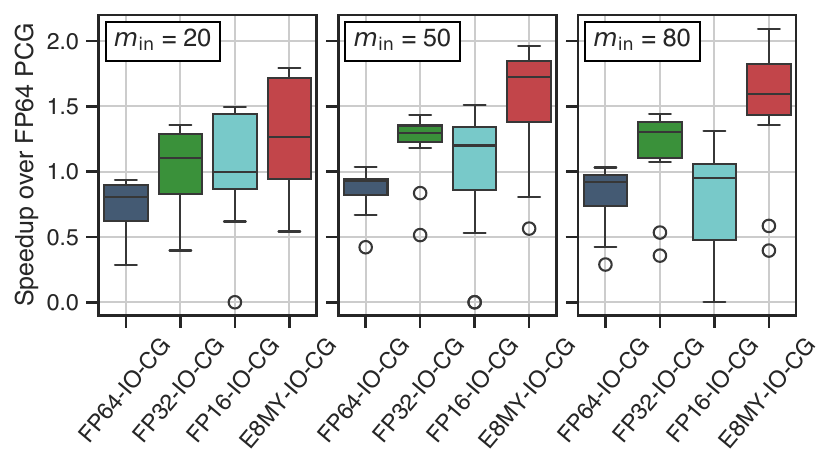}
  \caption{Performance of four mixed-precision inner-outer CG variants relative to the standard FP64 PCG solver.}
  \label{fig:cg}
\end{figure}

\begin{table}[t]
  \caption{Formats That Achieved the Best Performance}
  \label{tab:iocg-setting}
  \centering
  \footnotesize
  \setlength{\tabcolsep}{5pt}
  \begin{tabular}{lrrr}
    \hline
    Matrix             & $m_\mathrm{in} = 20$ & $m_\mathrm{in} = 50$ & $m_\mathrm{in} = 80$ \\
    \hline
    \mname{Bump_2911}  & E8M11                & E8M14                & E8M15                \\
    \mname{Emilia_923} & E8M12                & E8M15                & E8M19                \\
    \mname{G3_circuit} & E8M14                & E8M13                & E8M17                \\
    \mname{Queen_4147} & E8M13                & E8M14                & E8M14                \\
    \mname{Serena}     & E8M13                & E8M12                & E8M11                \\
    \mname{apache2}    & E8M20                & E8M15                & E8M21                \\
    \mname{audikw_1}   & E8M13                & E8M14                & E8M16                \\
    \mname{ecology2}   & E8M11                & E8M12                & E8M12                \\
    \mname{ldoor}      & E8M13                & E8M20                & E8M17                \\
    \mname{thermal2}   & E8M11                & E8M16                & E8M16                \\
    \mname{tmt_sym}    & E8M10                & E8M10                & E8M20                \\
    \mname{HPCG_7_7_7} & E8M10                & E8M11                & E8M14                \\
    \mname{HPCG_8_7_7} & E8M10                & E8M11                & E8M14                \\
    \mname{HPCG_8_8_7} & E8M11                & E8M13                & E8M13                \\
    \mname{HPCG_8_8_8} & E8M10                & E8M12                & E8M14                \\
    \hline
  \end{tabular}
\end{table}

We applied the SD-AINV preconditioner to all solvers, and tested three settings for the number of inner iterations $m_\mathrm{in} = 20$, 50, and 80.
\Cref{fig:cg} shows the achieved performance of the four IO-CG variants relative to the standard FP64 PCG solver;
for E8M$Y$-IO-CG, it reports the best results obtained by selecting the optimal E8M$Y$ format listed in \cref{tab:iocg-setting}.

First, FP64-IO-CG generally underperformed FP64 PCG due to the increased iterations and operations introduced by the inner-outer scheme.
However, this overhead became less significant as $m_\mathrm{in}$ increased;
for $m_\mathrm{in}=50$ and 80, FP64-IO-CG exhibited convergence behavior and performance close to PCG.
Due to these properties, FP32-IO-CG outperformed PCG in many cases by reducing data movement.
Nevertheless, its speedup over PCG remained around $1.3\times$, which is not particularly remarkable compared to the ideal gain and the speedups achieved in previous studies~\citep{buttari2008UsingMixed,yamazaki2022MixedPrecision,guo2025AdaptiveMixed,chen2026Enablingmixedprecision}.

\begin{figure}[tbp]
  \centering
  \subfloat[\mname{HPCG_8_8_8}\label{sfig:hpcg}]{
    \includegraphics[width=0.7\linewidth]{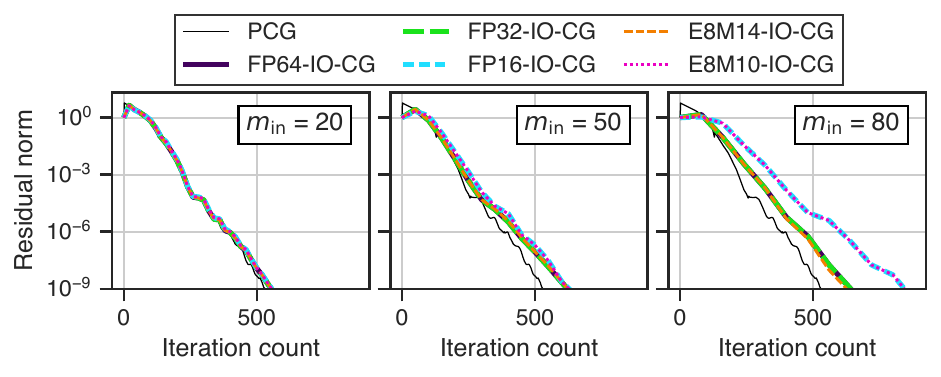}}\\
  \vspace{-2mm}
  \subfloat[\mname{ldoor}\label{sfig:ldoor}]{
    \includegraphics[width=0.7\linewidth]{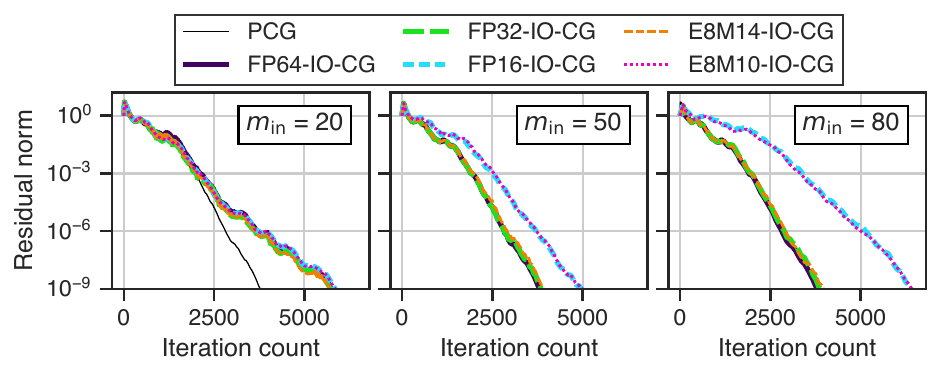}}
  \caption{History of the relative residual norm. For IO-CG, the iteration count denotes the number of inner iterations.}
  \label{fig:cg-case1}
\end{figure}

One approach to further improve solver performance is to use FP16.
However, as shown in \cref{fig:cg-case1}, larger $m_\mathrm{in}$ also degraded convergence, particularly in FP16-IO-CG, because the precision became insufficient during many inner iterations.
Even when FP32-IO-CG behaved similarly to FP64-IO-CG, FP16-IO-CG required more iterations (e.g., for \mname{ldoor});
this increase outweighed the benefits of reduced data movement, resulting in performance similar to or inferior to FP32-IO-CG.
A well-known non-IEEE alternative is BF16 (E8M7).
In our experiments, however, E8M10 showed convergence similar to FP16 (\cref{fig:cg-case1}).
This indicates that the number of mantissa bits, rather than exponent bits, is critical for good convergence and that BF16 is also ineffective in this context.

By contrast, as discussed in \cref{ssub:spmv_for_various_formats}, \proposed{} can control the mantissa length of E8M$Y$, achieving FP32-level accuracy while outperforming both FP32 and FP16 SpMV kernels.
Owing to these advantages, E8M$Y$-IO-CG outperformed FP64 PCG on 13 of the 15 problems, even for $m_\mathrm{in} = 80$.
Indeed, \cref{fig:cg-case1} shows that its convergence behavior closely matched that of FP32-IO-CG, for instance at $Y = 14$.
Consequently, unlike FP32-IO-CG and FP16-IO-CG, E8M$Y$-IO-CG attained speedups of up to 2.0$\times$ relative to FP64 PCG, which is a notable improvement over previous techniques and approaching the ideal gain.
Furthermore, \cref{tab:iocg-setting} shows that the best format tended to require more mantissa bits as $m_\mathrm{in}$ increased, in order to maintain sufficient accuracy over many inner iterations.
That is, the ability of \proposed{} to finely tune the mantissa length benefits the inner-outer scheme in IO-CG.

Admittedly, IO-CG requires more memory than PCG, and several practical considerations remain.
However, these results demonstrate that \proposed{} can realize solvers that outperform PCG with gains close to the ideal, which are difficult to achieve when relying on standard formats such as FP16 and BF16 in this IO-CG scheme.
Therefore, \proposed{} and its core idea of packing value and index information to enable flexible data representation are expected to be a useful basis for future research on advanced mixed-precision solvers, including more practical variants of IO-CG.

\section{Conclusions}
We propose a new sparse matrix format, \proposed{}, based on SELL, for high-performance SpMV on GPUs across various data representations.
It incorporates delta encoding of column indices and a new delta–value packing scheme to reduce memory footprint and data movement.
This packing scheme also allows flexible data representations by adjusting the bit allocation between deltas and values.

The numerical results on an NVIDIA A100 GPU show that SELL is effective for 292 of the 435 practical matrices tested, indicating that improving SELL-based SpMV is meaningful.
Although hardware-oriented optimizations particularly for smaller matrices remain future work, \proposed{} performs well across a variety of data representations, including custom non-IEEE formats.
For example, on 201 of these 292 matrices, it outperforms the cuSPARSE SELL kernel in FP16 SpMV by up to $1.63\times$.

Furthermore, evaluations with Krylov subspace methods confirm its practical effectiveness.
In the FP16-enabled F3R solver, using \proposed{} for FP16 SpMV improves the overall performance by up to 1.32$\times$.
\proposed{} is also effective when FP32 is costly and FP16 lacks sufficient accuracy, due to its ability to support custom data formats.
An inner-outer CG method using \proposed{} with the E8M$Y$ format outperforms FP64 PCG by up to $2.09\times$.
Mixed-precision solvers using non-IEEE formats remain relatively underexplored because of the limited support of these formats in high-performance SpMV kernels.
Thus, \proposed{} not only improves FP16-enabled solvers, but also provides a practical means to further investigate advanced mixed-precision solvers that utilize custom data representations.

Future work includes three directions.
First, evaluating \proposed{}-based SpMV on other hardware platforms may reveal different performance trends.
While a similar behavior is expected in memory-bound scenarios, other hardware features such as cache size may change the importance of hardware-oriented optimizations.
Second, extending the delta–value packing scheme to CSR-based formats may address the cases where SELL needs excessive padding, though this may require careful handling of load balancing and inter-thread communication.
Third, applying \proposed{} to other sparse matrix kernels, such as sparse triangular solves, is promising because some of their implementations are similar to SpMV.

\section*{Acknowledgment}
This work was supported by JSPS KAKENHI Grant Number JP25K24388.

\bibliographystyle{ACM-Reference-Format}
\bibliography{pep-references}

@article{abdelfattah2021surveynumerical,
  title = {A Survey of Numerical Linear Algebra Methods Utilizing Mixed-Precision Arithmetic},
  author = {Abdelfattah, Ahmad and Anzt, Hartwig and Boman, Erik G and Carson, Erin and Cojean, Terry and Dongarra, Jack and Fox, Alyson and Gates, Mark and Higham, Nicholas J and Li, Xiaoye S and Loe, Jennifer and Luszczek, Piotr and Pranesh, Srikara and Rajamanickam, Siva and Ribizel, Tobias and Smith, Barry F and Swirydowicz, Kasia and Thomas, Stephen and Tomov, Stanimire and Tsai, Yaohung M and Yang, Ulrike Meier},
  year = 2021,
  month = jul,
  journal = {The International Journal of High Performance Computing Applications},
  volume = {35},
  number = {4},
  pages = {344--369},
  doi = {10.1177/10943420211003313},
  url = {https://journals.sagepub.com/doi/10.1177/10943420211003313},
  langid = {english}
}

@article{aliaga2022Compressionload,
  title = {Compression and Load Balancing for Efficient Sparse Matrix-vector Product on Multicore Processors and Graphics Processing Units},
  author = {Aliaga, Jos{\'e} I. and Anzt, Hartwig and Gr{\"u}tzmacher, Thomas and Quintana-Ort{\'i}, Enrique S. and Tom{\'a}s, Andr{\'e}s E.},
  year = 2022,
  month = jun,
  journal = {Concurrency and Computation},
  volume = {34},
  number = {14},
  pages = {e6515},
  doi = {10.1002/cpe.6515},
  url = {https://onlinelibrary.wiley.com/doi/10.1002/cpe.6515},
  langid = {english}
}

@article{amestoy2024FivePrecisionGMRESBased,
  title = {Five-{{Precision GMRES-Based Iterative Refinement}}},
  author = {Amestoy, Patrick and Buttari, Alfredo and Higham, Nicholas J. and L'Excellent, Jean-Yves and Mary, Theo and Vieubl{\'e}, Bastien},
  year = 2024,
  month = mar,
  journal = {SIAM J. Matrix Anal. Appl.},
  volume = {45},
  number = {1},
  pages = {529--552},
  doi = {10.1137/23M1549079},
  url = {https://epubs.siam.org/doi/10.1137/23M1549079},
  langid = {english}
}

@inproceedings{anderson2016VectorizationMultibyte,
  title = {Vectorization of {{Multibyte Floating Point Data Formats}}},
  booktitle = {Proc. 2016 {{Int}}. {{Conf}}. {{Parallel Archit}}. {{Compil}}.},
  author = {Anderson, Andrew and Gregg, David},
  year = 2016,
  month = sep,
  series = {{{PACT}} '16},
  pages = {363--372},
  publisher = {Association for Computing Machinery},
  address = {New York, NY, USA},
  doi = {10.1145/2967938.2967966},
  url = {https://dl.acm.org/doi/10.1145/2967938.2967966}
}

@techreport{anzt2014ImplementingSparse,
  title = {Implementing a {{Sparse Matrix Vector Product}} for the {{SELL-C}}/{{SELL-C-$\sigma$}} Formats on {{NVIDIA GPUs}}},
  author = {Anzt, Hartwig and Tomov, Stanimire and Dongarra, Jack},
  year = 2014,
  institution = {University of Tennessee},
  langid = {english}
}

@article{anzt2020LoadbalancingSparse,
  title = {Load-Balancing {{Sparse Matrix Vector Product Kernels}} on {{GPUs}}},
  author = {Anzt, Hartwig and Cojean, Terry and {Yen-Chen}, Chen and Dongarra, Jack and Flegar, Goran and Nayak, Pratik and Tomov, Stanimire and Tsai, Yuhsiang M. and Wang, Weichung},
  year = 2020,
  month = mar,
  journal = {ACM Trans. Parallel Comput.},
  volume = {7},
  number = {1},
  pages = {1--26},
  doi = {10.1145/3380930},
  url = {https://dl.acm.org/doi/10.1145/3380930},
  langid = {english}
}

@inproceedings{ashari2014FastSparse,
  title = {Fast {{Sparse Matrix-Vector Multiplication}} on {{GPUs}} for {{Graph Applications}}},
  booktitle = {{{SC14 Int}}. {{Conf}}. {{High Perform}}. {{Comput}}. {{Netw}}. {{Storage Anal}}.},
  author = {Ashari, Arash and Sedaghati, Naser and Eisenlohr, John and Parthasarath, Srinivasan and Sadayappan, P.},
  year = 2014,
  month = nov,
  pages = {781--792},
  publisher = {IEEE},
  address = {New Orleans, LA, USA},
  doi = {10.1109/SC.2014.69},
  url = {http://ieeexplore.ieee.org/document/7013051/},
  langid = {english}
}

@book{barrett1994Templatessolution,
  title = {Templates for the Solution of Linear Systems: Building Blocks for Iterative Methods},
  author = {Barrett, Richard and Berry, Michael and Chan, Tony F. and Demmel, James and Donato, June and Dongarra, Jack and Eijkhout, Victor and Pozo, Roldan and Romine, Charles and van der Vorst, Henk},
  year = 1994,
  publisher = {SIAM, Philadelphia, PA},
  doi = {10.1137/1.9781611971538},
  url = {https://doi.org/10.1137/1.9781611971538}
}

@inproceedings{bell2009Implementingsparse,
  title = {Implementing Sparse Matrix-Vector Multiplication on Throughput-Oriented Processors},
  booktitle = {Proc. {{Conf}}. {{High Perform}}. {{Comput}}. {{Netw}}. {{Storage Anal}}.},
  author = {Bell, Nathan and Garland, Michael},
  year = 2009,
  month = nov,
  series = {{{SC}} '09},
  pages = {1--11},
  publisher = {Association for Computing Machinery},
  address = {New York, NY, USA},
  doi = {10.1145/1654059.1654078},
  url = {https://dl.acm.org/doi/10.1145/1654059.1654078}
}

@article{buttari2008UsingMixed,
  title = {Using {{Mixed Precision}} for {{Sparse Matrix Computations}} to {{Enhance}} the {{Performance}} While {{Achieving}} 64-Bit {{Accuracy}}},
  author = {Buttari, Alfredo and Dongarra, Jack and Kurzak, Jakub and Luszczek, Piotr and Tomov, Stanimir},
  year = 2008,
  month = jul,
  journal = {ACM Trans. Math. Softw.},
  volume = {34},
  number = {4},
  pages = {1--22},
  doi = {10.1145/1377596.1377597},
  url = {https://dl.acm.org/doi/10.1145/1377596.1377597},
  langid = {english}
}

@article{chen2026Enablingmixedprecision,
  title = {Enabling Mixed-Precision in Spectral Element Codes},
  author = {Chen, Yanxiang and De Oliveira Castro, Pablo and Bientinesi, Paolo and Jansson, Niclas and Iakymchuk, Roman},
  year = 2026,
  month = jan,
  journal = {Future Generation Computer Systems},
  volume = {174},
  pages = {107990},
  doi = {10.1016/j.future.2025.107990},
  url = {https://linkinghub.elsevier.com/retrieve/pii/S0167739X25002857},
  langid = {english}
}

@inproceedings{cong2025CBSpMVData,
  title = {{{CB-SpMV}}:{{A Data Aggregating}} and {{Balance Algorithm}} for for {{Cache-Friendly Block-Based SpMV}} on {{GPUs}}},
  shorttitle = {{{CB-SpMV}}},
  booktitle = {Proc. 39th {{ACM Int}}. {{Conf}}. {{Supercomput}}.},
  author = {Cong, Xing and Sun, FuKai and Chen, YiFan and Xie, Chenhao and Liu, Yi and Qian, Depei},
  year = 2025,
  month = jun,
  pages = {149--160},
  publisher = {ACM},
  address = {Salt Lake City USA},
  doi = {10.1145/3721145.3725746},
  url = {https://dl.acm.org/doi/10.1145/3721145.3725746},
  langid = {english}
}

@article{davis2011UniversityFlorida,
  title = {The {{University}} of {{Florida}} Sparse Matrix Collection},
  author = {Davis, Timothy A and Hu, Yifan},
  year = 2011,
  journal = {ACM Trans. Math. Softw.},
  volume = {38},
  number = {1},
  pages = {1--25},
  publisher = {ACM New York, NY, USA},
  doi = {10.1145/2049662.2049663}
}

@article{dongarra2016Highperformanceconjugategradient,
  title = {High-Performance Conjugate-Gradient Benchmark: {{A}} New Metric for Ranking High-Performance Computing Systems},
  shorttitle = {High-Performance Conjugate-Gradient Benchmark},
  author = {Dongarra, Jack and Heroux, Michael A and Luszczek, Piotr},
  year = 2016,
  month = feb,
  journal = {The International Journal of High Performance Computing Applications},
  volume = {30},
  number = {1},
  pages = {3--10},
  doi = {10.1177/1094342015593158},
  url = {https://journals.sagepub.com/doi/10.1177/1094342015593158},
  langid = {english}
}

@article{filippone2017SparseMatrixVector,
  title = {Sparse {{Matrix-Vector Multiplication}} on {{GPGPUs}}},
  author = {Filippone, Salvatore and Cardellini, Valeria and Barbieri, Davide and Fanfarillo, Alessandro},
  year = 2017,
  month = jan,
  journal = {ACM Trans. Math. Softw.},
  volume = {43},
  number = {4},
  pages = {30:1--30:49},
  doi = {10.1145/3017994},
  url = {https://dl.acm.org/doi/10.1145/3017994}
}

@inproceedings{galanopoulos2025DIVIndex,
  title = {{{DIV}}: {{An Index}} \& {{Value}} Compression Method for {{SpMV}} on Large Matrices},
  shorttitle = {{{DIV}}},
  booktitle = {Proc. 39th {{ACM Int}}. {{Conf}}. {{Supercomput}}.},
  author = {Galanopoulos, Dimitrios and Mpakos, Panagiotis and Anastasiadis, Petros and Koziris, Nectarios and Goumas, Georgios},
  year = 2025,
  month = jun,
  pages = {705--717},
  publisher = {ACM},
  address = {Salt Lake City USA},
  doi = {10.1145/3721145.3725767},
  url = {https://dl.acm.org/doi/10.1145/3721145.3725767},
  langid = {english}
}

@misc{gao2024SystematicLiterature,
  title = {A {{Systematic Literature Survey}} of {{Sparse Matrix-Vector Multiplication}}},
  author = {Gao, Jianhua and Liu, Bingjie and Ji, Weixing and Huang, Hua},
  year = 2024,
  month = apr,
  number = {arXiv:2404.06047},
  eprint = {2404.06047},
  primaryclass = {cs},
  publisher = {arXiv},
  doi = {10.48550/arXiv.2404.06047},
  url = {http://arxiv.org/abs/2404.06047},
  archiveprefix = {arXiv}
}

@article{graillat2024AdaptivePrecision,
  title = {Adaptive {{Precision Sparse Matrix}}--{{Vector Product}} and {{Its Application}} to {{Krylov Solvers}}},
  author = {Graillat, Stef and J{\'e}z{\'e}quel, Fabienne and Mary, Th{\'e}o and Molina, Rom{\'e}o},
  year = 2024,
  journal = {SIAM J. Sci. Comput.},
  volume = {46},
  number = {1},
  pages = {C30--C56},
  publisher = {SIAM},
  doi = {10.1137/22M1522619}
}

@incollection{graillat2024ReducedPrecisionReducedExponent,
  title = {Reduced-{{Precision}} and {{Reduced-Exponent Formats}} for {{Accelerating Adaptive Precision Sparse Matrix}}--{{Vector Product}}},
  booktitle = {Euro-{{Par}} 2024: {{Parallel Processing}}},
  author = {Graillat, Stef and J{\'e}z{\'e}quel, Fabienne and Mary, Theo and Molina, Rom{\'e}o and Mukunoki, Daichi},
  editor = {Carretero, Jesus and Shende, Sameer and {Garcia-Blas}, Javier and Brandic, Ivona and Olcoz, Katzalin and Schreiber, Martin},
  year = 2024,
  volume = {14803},
  pages = {17--30},
  publisher = {Springer Nature Switzerland},
  address = {Cham},
  doi = {10.1007/978-3-031-69583-4_2},
  url = {https://link.springer.com/10.1007/978-3-031-69583-4_2},
  langid = {english}
}

@misc{guo2025AdaptiveMixed,
  title = {An {{Adaptive Mixed Precision}} and {{Dynamically Scaled Preconditioned Conjugate Gradient Algorithm}}},
  author = {Guo, Yichen and de Sturler, Eric and Warburton, Tim},
  year = 2025,
  month = may,
  number = {arXiv:2505.04155},
  eprint = {2505.04155},
  primaryclass = {math},
  publisher = {arXiv},
  doi = {10.48550/arXiv.2505.04155},
  url = {http://arxiv.org/abs/2505.04155},
  archiveprefix = {arXiv}
}

@inproceedings{hunhold2025EvaluationBfloat16,
  title = {Evaluation of {{Bfloat16}}, {{Posit}}, and {{Takum Arithmetics}} in {{Sparse Linear Solvers}}},
  booktitle = {2025 {{IEEE}} 32nd {{Symp}}. {{Comput}}. {{Arith}}. {{ARITH}}},
  author = {Hunhold, Laslo and Quinlan, James},
  year = 2025,
  month = may,
  pages = {61--68},
  publisher = {IEEE},
  doi = {10.1109/ARITH64983.2025.00019},
  url = {https://ieeexplore.ieee.org/document/11038220/}
}

@inproceedings{ichimura2018FastScalable,
  title = {A {{Fast Scalable Implicit Solver}} for {{Nonlinear Time-Evolution Earthquake City Problem}} on {{Low-Ordered Unstructured Finite Elements}} with {{Artificial Intelligence}} and {{Transprecision Computing}}},
  booktitle = {{{SC18 Int}}. {{Conf}}. {{High Perform}}. {{Comput}}. {{Netw}}. {{Storage Anal}}.},
  author = {Ichimura, Tsuyoshi and Fujita, Kohei and Yamaguchi, Takuma and Naruse, Akira and Wells, Jack C. and Schulthess, Thomas C. and Straatsma, Tjerk P. and Zimmer, Christopher J. and Martinasso, Maxime and Nakajima, Kengo and Hori, Muneo and Maddegedara, Lalith},
  year = 2018,
  month = nov,
  pages = {627--637},
  publisher = {IEEE},
  address = {Dallas, TX, USA},
  doi = {10.1109/SC.2018.00052},
  url = {https://ieeexplore.ieee.org/document/8665778/},
  langid = {english}
}

@article{ikuno2012IterativeSolver,
  title = {Iterative {{Solver}} for {{Linear System Obtained}} by {{Edge Element}}: {{Variable Preconditioned Method With Mixed Precision}} on {{GPU}}},
  shorttitle = {Iterative {{Solver}} for {{Linear System Obtained}} by {{Edge Element}}},
  author = {Ikuno, Soichiro and Kawaguchi, Yuki and Fujita, Norihisa and Itoh, Taku and Nakata, Susumu and Watanabe, Kota},
  year = 2012,
  month = feb,
  journal = {IEEE Trans. Magn.},
  volume = {48},
  number = {2},
  pages = {467--470},
  doi = {10.1109/TMAG.2011.2175375},
  url = {http://ieeexplore.ieee.org/document/6136645/},
  copyright = {https://ieeexplore.ieee.org/Xplorehelp/downloads/license-information/IEEE.html},
  langid = {english}
}

@inproceedings{iwashita2020integerarithmeticbased,
  title = {An Integer Arithmetic-Based Sparse Linear Solver Using a {{GMRES}} Method and Iterative Refinement},
  booktitle = {2020 {{IEEEACM}} 11th {{Workshop Latest Adv}}. {{Scalable Algorithms Large-Scale Syst}}. {{ScalA}}},
  author = {Iwashita, Takeshi and Suzuki, Kengo and Fukaya, Takeshi},
  year = 2020,
  pages = {1--8},
  publisher = {IEEE},
  doi = {10.1109/ScalA51936.2020.00006}
}

@article{jerez2015LowComplexity,
  title = {A {{Low Complexity Scaling Method}} for the {{Lanczos Kernel}} in {{Fixed-Point Arithmetic}}},
  author = {Jerez, Juan Luis and Constantinides, George A. and Kerrigan, Eric C.},
  year = 2015,
  journal = {IEEE Trans. Comput.},
  volume = {64},
  number = {2},
  pages = {303--315},
  doi = {10.1109/TC.2013.162}
}

@inproceedings{kawai2022LowAdaptive,
  title = {Low/{{Adaptive Precision Computation}} in {{Preconditioned Iterative Solvers}} for {{Ill-Conditioned Problems}}},
  booktitle = {Int. {{Conf}}. {{High Perform}}. {{Comput}}. {{Asia-Pac}}. {{Reg}}.},
  author = {Kawai, Masatoshi and Nakajima, Kengo},
  year = 2022,
  month = jan,
  series = {{{HPCAsia}} '22},
  pages = {30--40},
  publisher = {Association for Computing Machinery},
  address = {New York, NY, USA},
  doi = {10.1145/3492805.3492813},
  url = {https://dl.acm.org/doi/10.1145/3492805.3492813}
}

@article{kreutzer2014unifiedsparse,
  title = {A Unified Sparse Matrix Data Format for Efficient General Sparse Matrix-Vector Multiply on Modern Processors with Wide {{SIMD}} Units},
  author = {Kreutzer, Moritz and Hager, Georg and Wellein, Gerhard and Fehske, Holger and Bishop, Alan R.},
  year = 2014,
  month = jan,
  journal = {SIAM J. Sci. Comput.},
  volume = {36},
  number = {5},
  eprint = {1307.6209},
  primaryclass = {cs},
  pages = {C401-C423},
  doi = {10.1137/130930352},
  url = {http://arxiv.org/abs/1307.6209},
  archiveprefix = {arXiv}
}

@article{lindquist2022AcceleratingRestarted,
  title = {Accelerating {{Restarted GMRES With Mixed Precision Arithmetic}}},
  author = {Lindquist, Neil and Luszczek, Piotr and Dongarra, Jack},
  year = 2022,
  month = apr,
  journal = {IEEE Trans. Parallel Distrib. Syst.},
  volume = {33},
  number = {4},
  pages = {1027--1037},
  doi = {10.1109/TPDS.2021.3090757},
  url = {https://ieeexplore.ieee.org/document/9462418/},
  langid = {english}
}

@phdthesis{lindquistReducingCommunication,
  title = {Reducing {{Communication}} in the {{Solution}} of {{Linear Systems}}},
  author = {Lindquist, Neil},
  year = 2023,
  month = aug,
  school = {The University of Tennessee, Knoxville},
  address = {TN, USA},
  langid = {english},
  url = {https://trace.tennessee.edu/handle/20.500.14382/29814}
}

@inproceedings{liu2015CSR5Efficient,
  title = {{{CSR5}}: {{An Efficient Storage Format}} for {{Cross-Platform Sparse Matrix-Vector Multiplication}}},
  shorttitle = {{{CSR5}}},
  booktitle = {Proc. 29th {{ACM Int}}. {{Conf}}. {{Supercomput}}.},
  author = {Liu, Weifeng and Vinter, Brian},
  year = 2015,
  month = jun,
  series = {{{ICS}} '15},
  pages = {339--350},
  publisher = {Association for Computing Machinery},
  address = {New York, NY, USA},
  doi = {10.1145/2751205.2751209},
  url = {https://dl.acm.org/doi/10.1145/2751205.2751209}
}

@inproceedings{lu2023DASPSpecific,
  title = {{{DASP}}: {{Specific Dense Matrix Multiply-Accumulate Units Accelerated General Sparse Matrix-Vector Multiplication}}},
  shorttitle = {{{DASP}}},
  booktitle = {Proc. {{Int}}. {{Conf}}. {{High Perform}}. {{Comput}}. {{Netw}}. {{Storage Anal}}.},
  author = {Lu, Yuechen and Liu, Weifeng},
  year = 2023,
  month = nov,
  pages = {1--14},
  publisher = {ACM},
  address = {Denver CO USA},
  doi = {10.1145/3581784.3607051},
  url = {https://dl.acm.org/doi/10.1145/3581784.3607051},
  langid = {english}
}

@inproceedings{maggioni2014CoAdELLAdaptivity,
  title = {{{CoAdELL}}: {{Adaptivity}} and {{Compression}} for {{Improving Sparse Matrix-Vector Multiplication}} on {{GPUs}}},
  shorttitle = {{{CoAdELL}}},
  booktitle = {2014 {{IEEE Int}}. {{Parallel Distrib}}. {{Process}}. {{Symp}}. {{Workshop}}},
  author = {Maggioni, Marco and {Berger-Wolf}, Tanya},
  year = 2014,
  month = may,
  pages = {933--940},
  publisher = {IEEE},
  doi = {10.1109/IPDPSW.2014.106},
  url = {https://ieeexplore.ieee.org/document/6969482/}
}

@incollection{monakov2010AutomaticallyTuninga,
  title = {Automatically {{Tuning Sparse Matrix-Vector Multiplication}} for {{GPU Architectures}}},
  booktitle = {High {{Performance Embedded Architectures}} and {{Compilers}}},
  author = {Monakov, Alexander and Lokhmotov, Anton and Avetisyan, Arutyun},
  editor = {Hutchison, David and Kanade, Takeo and Kittler, Josef and Kleinberg, Jon M. and Mattern, Friedemann and Mitchell, John C. and Naor, Moni and Nierstrasz, Oscar and Pandu Rangan, C. and Steffen, Bernhard and Sudan, Madhu and Terzopoulos, Demetri and Tygar, Doug and Vardi, Moshe Y. and Weikum, Gerhard and Patt, Yale N. and Foglia, Pierfrancesco and Duesterwald, Evelyn and Faraboschi, Paolo and Martorell, Xavier},
  year = 2010,
  volume = {5952},
  pages = {111--125},
  publisher = {Springer Berlin Heidelberg},
  address = {Berlin, Heidelberg},
  doi = {10.1007/978-3-642-11515-8_10},
  url = {http://link.springer.com/10.1007/978-3-642-11515-8_10},
  langid = {english}
}

@inproceedings{mukunoki2023SparseMatrixVector,
  title = {Sparse {{Matrix-Vector Multiplication}} with {{Reduced-Precision Memory Accessor}}},
  booktitle = {2023 {{IEEE}} 16th {{Int}}. {{Symp}}. {{Embed}}. {{MulticoreMany-Core Syst}}.--{{Chip MCSoC}}},
  author = {Mukunoki, Daichi and Kawai, Masatoshi and Imamura, Toshiyuki},
  year = 2023,
  month = dec,
  pages = {608--615},
  publisher = {IEEE},
  doi = {10.1109/MCSoC60832.2023.00094},
  url = {https://ieeexplore.ieee.org/document/10387875/}
}

@article{murakami2026CoDSELLNonZero,
  title = {{{CoD-SELL}}: {{A Non-Zero Location Dictionary Compression Sparse Matrix Format}} for {{SpMV}} on {{GPU}}},
  shorttitle = {{{CoD-SELL}}},
  author = {Murakami, Shun and Yoneda, Kazunori and Iwamura, Takashi and Watanabe, Masahiro and Inoguchi, Yasushi},
  year = 2026,
  journal = {IEEE Access},
  volume = {14},
  pages = {17058--17068},
  doi = {10.1109/ACCESS.2026.3659140},
  url = {https://ieeexplore.ieee.org/document/11367640/}
}

@article{nagasaka2016AdaptiveMultilevel,
  title = {Adaptive {{Multi-level Blocking Optimization}} for {{Sparse Matrix Vector Multiplication}} on {{GPU}}},
  author = {Nagasaka, Yusuke and Nukada, Akira and Matsuoka, Satoshi},
  year = 2016,
  journal = {Procedia Computer Science},
  volume = {80},
  pages = {131--142},
  doi = {10.1016/j.procs.2016.05.304},
  url = {https://linkinghub.elsevier.com/retrieve/pii/S187705091630655X},
  langid = {english}
}

@inproceedings{nakajima2021EfficientParallel,
  title = {Efficient {{Parallel Multigrid Methods}} on {{Manycore Clusters}} with {{Double}}/{{Single Precision Computing}}},
  booktitle = {2021 {{IEEE Int}}. {{Parallel Distrib}}. {{Process}}. {{Symp}}. {{Workshop IPDPSW}}},
  author = {Nakajima, Kengo and Ogita, Takseshi and Kawai, Masatoshi},
  year = 2021,
  month = jun,
  pages = {760--769},
  publisher = {IEEE},
  doi = {10.1109/IPDPSW52791.2021.00114},
  url = {https://ieeexplore.ieee.org/document/9460632/}
}

@inproceedings{niu2021TileSpMVTiled,
  title = {{{TileSpMV}}: {{A Tiled Algorithm}} for {{Sparse Matrix-Vector Multiplication}} on {{GPUs}}},
  shorttitle = {{{TileSpMV}}},
  booktitle = {2021 {{IEEE Int}}. {{Parallel Distrib}}. {{Process}}. {{Symp}}. {{IPDPS}}},
  author = {Niu, Yuyao and Lu, Zhengyang and Dong, Meichen and Jin, Zhou and Liu, Weifeng and Tan, Guangming},
  year = 2021,
  month = may,
  pages = {68--78},
  publisher = {IEEE},
  doi = {10.1109/IPDPS49936.2021.00016},
  url = {https://ieeexplore.ieee.org/document/9460505/}
}

@article{notay2000FlexibleConjugate,
  title = {Flexible {{Conjugate Gradients}}},
  author = {Notay, Yvan},
  year = 2000,
  month = jan,
  journal = {SIAM J. Sci. Comput.},
  volume = {22},
  number = {4},
  pages = {1444--1460},
  doi = {10.1137/S1064827599362314},
  url = {http://epubs.siam.org/doi/10.1137/S1064827599362314},
  langid = {english}
}

@article{spyropoulos2025Numericalstudy,
  title = {Numerical Study of Mixed Precision {{GMRES}}(m) Preconditioned by Deflation},
  author = {Spyropoulos, Antony and Antonopoulos, Christos},
  year = 2025,
  month = may,
  journal = {Numer Algor},
  volume = {101},
  pages = {2631--2657},
  doi = {10.1007/s11075-025-02102-z},
  url = {https://link.springer.com/10.1007/s11075-025-02102-z},
  langid = {english}
}

@inproceedings{steinberger2017Globallyhomogeneous,
  title = {Globally Homogeneous, Locally Adaptive Sparse Matrix-Vector Multiplication on the {{GPU}}},
  booktitle = {Proc. {{Int}}. {{Conf}}. {{Supercomput}}.},
  author = {Steinberger, Markus and Zayer, Rhaleb and Seidel, Hans-Peter},
  year = 2017,
  month = jun,
  series = {{{ICS}} '17},
  pages = {1--11},
  publisher = {Association for Computing Machinery},
  address = {New York, NY, USA},
  doi = {10.1145/3079079.3079086},
  url = {https://dl.acm.org/doi/10.1145/3079079.3079086}
}

@article{suzuki2022NewAINV,
  title = {A {{New AINV Preconditioner}} for the {{CG Method}} in {{Hybrid CPU-GPU Computing Environment}}},
  author = {Suzuki, Kengo and Fukaya, Takeshi and Iwashita, Takeshi},
  year = 2022,
  journal = {Journal of Information Processing},
  volume = {30},
  number = {0},
  pages = {755--765},
  doi = {10.2197/ipsjjip.30.755},
  url = {https://www.jstage.jst.go.jp/article/ipsjjip/30/0/30_755/_article},
  langid = {english}
}

@article{suzuki2025IntegerArithmeticBased,
  title = {An {{Integer Arithmetic-Based AMG Preconditioned FGMRES Solver}}},
  author = {Suzuki, Kengo and Fukaya, Takeshi and Iwashita, Takeshi},
  year = 2025,
  month = mar,
  journal = {ACM Trans. Math. Softw.},
  volume = {51},
  number = {1},
  pages = {1:1--1:25},
  doi = {10.1145/3704726},
  url = {https://dl.acm.org/doi/10.1145/3704726}
}

@inproceedings{suzuki2025NestedKrylov,
  title = {A {{Nested Krylov Method Using Half-Precision Arithmetic}}},
  booktitle = {Proc. {{Int}}. {{Conf}}. {{High Perform}}. {{Comput}}. {{Netw}}. {{Storage Anal}}.},
  author = {Suzuki, Kengo and Iwashita, Takeshi},
  year = 2025,
  month = nov,
  series = {{{SC}} '25},
  pages = {711--727},
  publisher = {Association for Computing Machinery},
  address = {New York, NY, USA},
  doi = {10.1145/3712285.3759807},
  url = {https://dl.acm.org/doi/10.1145/3712285.3759807}
}

@inproceedings{wolfgang2024ValueCompressedSparse,
  title = {Value-{{Compressed Sparse Column}} ({{VCSC}}): {{Sparse Matrix Storage}} for {{Single-cell Omics Data}}},
  shorttitle = {Value-{{Compressed Sparse Column}} ({{VCSC}})},
  booktitle = {2024 {{IEEE Int}}. {{Conf}}. {{Big Data BigData}}},
  author = {Wolfgang, Seth and Ruiter, Skyler and Tunnell, Marc and Triche, Timothy and Carrier, Erin and DeBruine, Zachary},
  year = 2024,
  month = dec,
  pages = {4952--4958},
  publisher = {IEEE},
  address = {Washington, DC, USA},
  doi = {10.1109/BigData62323.2024.10825091},
  url = {https://ieeexplore.ieee.org/document/10825091/},
  copyright = {https://doi.org/10.15223/policy-029},
  langid = {english}
}

@inproceedings{yamazaki2022HighPerformanceGMRES,
  title = {High-{{Performance GMRES Multi-Precision Benchmark}}: {{Design}}, {{Performance}}, and {{Challenges}}},
  booktitle = {2022 {{IEEEACM Int}}. {{Workshop Perform}}. {{Model}}. {{Benchmarking Simul}}. {{High Perform}}. {{Comput}}. {{Syst}}. {{PMBS}}},
  author = {Yamazaki, Ichitaro and Glusa, Christian and Loe, Jennifer and Luszczek, Piotr and Rajamanickam, Sivasankaran and Dongarra, Jack},
  year = 2022,
  pages = {112--122},
  publisher = {IEEE},
  doi = {10.1109/PMBS56514.2022.00015}
}

@inproceedings{yamazaki2022MixedPrecision,
  title = {Mixed {{Precision}} S-Step {{Conjugate Gradient}} with {{Residual Replacement}} on {{GPUs}}},
  booktitle = {2022 {{IEEE Int}}. {{Parallel Distrib}}. {{Process}}. {{Symp}}. {{IPDPS}}},
  author = {Yamazaki, Ichitaro and Carson, Erin and Kelley, Brian},
  year = 2022,
  month = may,
  pages = {886--896},
  publisher = {IEEE},
  doi = {10.1109/IPDPS53621.2022.00091},
  url = {https://ieeexplore.ieee.org/document/9820637/}
}

@inproceedings{zhang2025CanTensor,
  title = {Can {{Tensor Cores Benefit Memory-Bound Kernels}}? ({{NO}}!)},
  shorttitle = {Can {{Tensor Cores Benefit Memory-Bound Kernels}}?},
  booktitle = {Proc. 17th {{Workshop Gen}}. {{Purp}}. {{Process}}. {{Using GPU}}},
  author = {Zhang, Lingqi and Huang, Jiajun and Di, Sheng and Matsuoka, Satoshi and Wahib, Mohamed},
  year = 2025,
  month = may,
  series = {{{GPGPU}} '25},
  pages = {28--34},
  publisher = {Association for Computing Machinery},
  address = {New York, NY, USA},
  doi = {10.1145/3725798.3725803},
  url = {https://dl.acm.org/doi/10.1145/3725798.3725803}
}

@article{zhao2022NumericalInvestigation,
  title = {Numerical {{Investigation}} into the {{Mixed Precision GMRES}}({\emph{m}}) {{Method Using FP64}} and {{FP32}}},
  author = {Zhao, Yingqi and Fukaya, Takeshi and Zhang, Linjie and Iwashita, Takeshi},
  year = 2022,
  journal = {J. Inf. Process.},
  volume = {30},
  pages = {525--537},
  doi = {10.2197/ipsjjip.30.525}
}

@article{zhao2023NumericalBehavior,
  title = {Numerical {{Behavior}} of {{Mixed Precision Iterative Refinement Using}} the {{BiCGSTAB Method}}},
  author = {Zhao, Yingqi and Fukaya, Takeshi and Iwashita, Takeshi},
  year = 2023,
  journal = {Journal of Information Processing},
  volume = {31},
  pages = {860--874},
  doi = {10.2197/ipsjjip.31.860},
  url = {https://www.jstage.jst.go.jp/article/ipsjjip/31/0/31_860/_article},
  langid = {english}
}

@software{kengo_suzuki_2025_16882405,
  author       = {Kengo Suzuki},
  title        = {suzuki-hpc/F3R: v1.0.2},
  month        = aug,
  year         = 2025,
  publisher    = {Zenodo},
  version      = {v1.0.2},
  doi          = {10.5281/zenodo.16882405},
}

\end{document}